\documentclass[aps,pra,superscriptaddress]{revtex4}

\usepackage{times}
\usepackage[pdftex]{graphicx}
\usepackage[pdftex]{epsfig}
\usepackage{amsmath}
\usepackage{pslatex}
\usepackage{amssymb}
\usepackage{amsfonts}
\usepackage{color}
\usepackage{wrapfig}
\usepackage{eucal}
\usepackage{hhline}
\usepackage{threeparttable}
\usepackage{supertabular}
\usepackage{multirow}
\usepackage{tabularx}

\newcolumntype{Y}{>{\centering\arraybackslash}X}

\def\be{\begin{equation}}
\def\ee{\end{equation}}
\def\bea{\begin{eqnarray}}
\def\eea{\end{eqnarray}}

\newcommand{\avg}[1]{\mbox{$\langle#1\rangle$}}

\newcommand{\opdagger}[2]{\mbox{$\hat{#1}_{#2}^{\dagger}$}}
\newcommand{\op}[2]{\mbox{$\hat{#1}_{#2}$}}

\def\omegap{\omega_\text{l}}
\def\tpd{\Delta}

\begin{document}

\pagenumbering{arabic}

\title{Laser cooling of a nanomechanical oscillator into its quantum ground state}

\author{Jasper Chan}
\author{T. P. Mayer Alegre}
\author{Amir H. Safavi-Naeini}
\author{Jeff T. Hill}
\author{Alex Krause}
\affiliation{Thomas J. Watson, Sr., Laboratory of Applied Physics, California Institute of Technology, Pasadena, CA 91125}
\author{Simon Gr\"oblacher}
\affiliation{Thomas J. Watson, Sr., Laboratory of Applied Physics, California Institute of Technology, Pasadena, CA 91125}
\affiliation{Vienna Center for Quantum Science and Technology (VCQ), Faculty of Physics, University of Vienna, Boltzmanngasse 5, A-1090 Vienna, Austria}
\author{Markus Aspelmeyer}
\affiliation{Vienna Center for Quantum Science and Technology (VCQ), Faculty of Physics, University of Vienna, Boltzmanngasse 5, A-1090 Vienna, Austria}
\author{Oskar Painter}
\email{opainter@caltech.edu}
\homepage{http://copilot.caltech.edu}
\affiliation{Thomas J. Watson, Sr., Laboratory of Applied Physics, California Institute of Technology, Pasadena, CA 91125}

\date{\today}

\begin{abstract}
  A patterned Si nanobeam is formed which supports co-localized acoustic and optical resonances that are coupled via radiation pressure.  Starting from a bath temperature of $T_b \approx 20$~K, the $3.68$~GHz nanomechanical mode is cooled into its quantum mechanical ground state utilizing optical radiation pressure.  The mechanical mode displacement fluctuations, imprinted on the transmitted cooling laser beam, indicate that a final phonon mode occupancy of $\bar{n}=0.85\pm0.04$ is obtained.  
\end{abstract}

\maketitle

The simple mechanical oscillator, canonically consisting of a coupled mass-spring system, is used in a wide variety of sensitive measurements, including the detection of weak forces~\cite{Braginsky1977} and small masses~\cite{Jensen2008}.  A classical oscillator can take on a well-defined amplitude of sinusoidal motion.  A quantum oscillator, however, has a lowest energy state, or ground state, with a finite amplitude uncertainty corresponding to the zero-point motion.  In our everyday experience mechanical oscillators are filled with many energy quanta due to interactions with their highly fluctuating thermal environment, and the oscillator's quantum nature is all but hidden.  Recently, in experiments performed at temperatures of a few hundredths of a Kelvin, engineered nanomechanical resonators coupled to electrical circuits have been measured to be oscillating quietly in their quantum ground state~\cite{OConnell2010,Teufel2011b}.  These experiments, in addition to providing a glimpse into the underlying quantum behavior of mesoscopic systems consisting of billions of atoms, represent the initial steps towards the use of mechanical elements as tools for quantum metrology~\cite{Caves1980a,Regal2008} or as a means to couple hybrid quantum systems~\cite{Wallquist2009,Stannigel2010,Safavi-Naeini2011a}.  In this work we have created a coupled, nanoscale optical and mechanical resonator~\cite{Eichenfield2009b} formed in a silicon microchip, in which radiation pressure from a laser is used to cool the mechanical motion down to the quantum ground state (average phonon occupancy, $\bar{n} = 0.85\pm0.04$).  Critically, this cooling is realized at an environmental temperature some thousand times larger than in previous experiments ($T_b \approx 20$~K), and paves the way for optical control of mesoscale mechanical oscillators in the quantum regime.

It has been known for some time~\cite{Cohen-Tannoudji1990} that atoms and ions nearly resonant with an applied laser beam (or series of beams) may be mechanically manipulated, even trapped and cooled down to the quantum ground state of their center-of-mass motion~\cite{Monroe1995}. Equally well known~\cite{Braginsky1977} has been the fact that radiation pressure can be exerted on regular dielectric (i.e., non-resonant) objects to damp and cool their mechanical motion. In so-called cavity-assisted schemes, the radiation pressure force is enhanced by coupling the motion of a mechanical object to the light field in an optical cavity. Pumping of the optical cavity by a single-frequency electromagnetic source  produces a coupling between the mechanical motion and the intensity of the electromagnetic field built-up in the resonator. As the radiation pressure force exerted on the mechanical object is proportional to the field intensity in the resonator, a form of dynamical back-action results~\cite{Braginsky1977,Kippenberg2008}. For a lower-frequency (red) detuning of the laser from the cavity, this leads to damping and cooling of the mechanical motion.

Recent experiments involving micro- and nanomechanical resonators, coupled to electromagnetic fields at optical and microwave frequencies, have demonstrated significant radiation pressure dynamic back-action~\cite{Kippenberg2008}. These structures have included Fabry-P\'erot cavities with mechanically-compliant miniature end-mirrors~\cite{Gigan2006,Arcizet2006b,Groeblacher2009a} or internal nanomembranes~\cite{Thompson2008}, whispering-gallery glass resonators~\cite{Riviere2010}, nanowires capacitively coupled to co-planar microwave transmission-line cavities~\cite{Regal2008,Rocheleau2010}, and lumped circuit microwave resonators with deformable, nanoscale, vacuum-gap capacitors~\cite{Teufel2011a}.  The first measurement of an engineered mesoscopic mechanical resonator predominantly in its quantum ground state, however, has been performed not using back-action cooling, but rather, using conventional cryogenic cooling (bath temperature $T_b\approx 25$~mK) of a high frequency, and thus lower thermal occupancy, oscillator~\cite{OConnell2010}.  Read-out and control of mechanical motion at the single quanta level was performed by strongly coupling the GHz-frequency piezoelectric mechanical resonator to a resonant superconducting quantum circuit.  Only recently have microwave systems, also operating at bath temperatures of $T_b \approx 25$~mK, utilized radiation pressure back-action to cool a high-$Q$, MHz-frequency mechanical oscillator to the ground state~\cite{Rocheleau2010,Teufel2011b}.                   

\begin{figure*}[t!]
\begin{center}
\includegraphics[width=0.9\columnwidth]{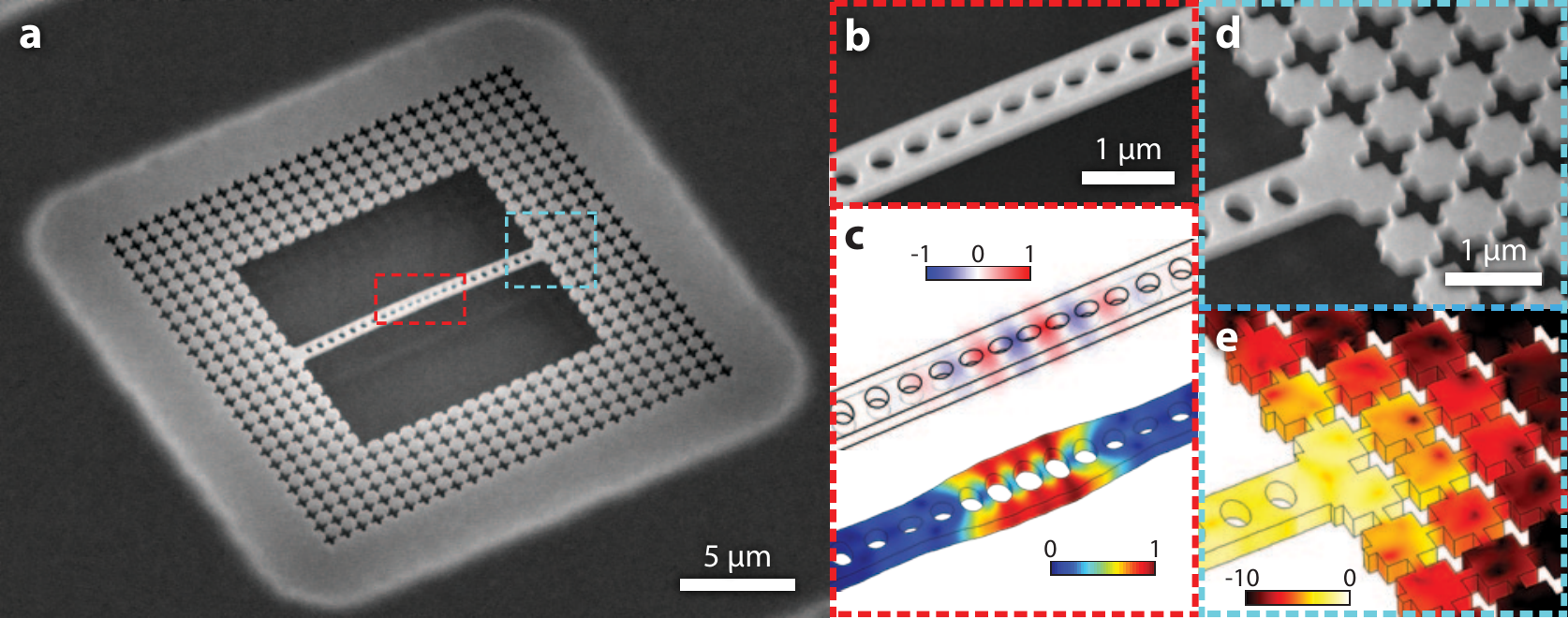}
\caption{\textbf{Optomechanical resonator with phononic shield.} \textbf{a}, Scanning electron microscope (SEM) image of the patterned Si nanobeam with external phononic bandgap shield. \textbf{b}, Enlarged SEM image of the central cavity region of the nanobeam. \textbf{c}, Top, finite element method (FEM) simulated normalized electric field of the localized optical resonance of the nanobeam cavity. Bottom, FEM simulated normalized displacement field of the acoustic resonance (breathing mode) which is coupled via radiation pressure to the co-localized optical resonance. The displacement field is indicated by the exaggerated deformation of the structure, with the relative magnitude of the local displacement (strain) indicated by the color. \textbf{d}, SEM image of the interface between the nanobeam and the phononic bandgap shield.  \textbf{e}, FEM simulation of the normalized squared displacement field amplitude of the localized acoustic resonance at the nanobeam-shield interface, indicating the strong suppression of acoustic radiation provided by the phononic bandgap shield. The color scale represents $\log(x^2/\max\{x^2\})$ where $x$ is the displacement.}
\label{fig:OM_device}
\end{center}
\end{figure*}

Optically coupled mechanical devices, while allowing for control of the mechanical system through well-established quantum optical techniques~\cite{Wiseman2010}, have thus far not reached the quantum regime due to a myriad of technical difficulties~\cite{Riviere2010}.  A particular challenge has been maintaining efficient optical coupling and low-loss optics and mechanics in a cryogenic, sub-Kelvin environment.  The optomechanical system studied in this work enables large optical coupling to a high-$Q$ GHz-frequency mechanical oscillator, allowing for both efficient back-action cooling and significantly higher operating temperatures.  As shown in Fig.~\ref{fig:OM_device}a, the system consists of an integrated optical and nanomechanical resonator formed in the surface layer of a silicon-on-insulator microchip. The periodic patterning of the nanobeam is designed to result in Bragg scattering of both optical and acoustic guided waves. A perturbation in the periodicity at the center of the beam results in co-localized optical and mechanical resonances (Fig.~\ref{fig:OM_device}b-c), which are coupled via radiation pressure~\cite{Eichenfield2009b}.  The fundamental optical resonance of the structure occurs at a frequency $\omega_o/2\pi = 195$~THz ($\lambda = 1537$~nm), while, due to the much slower speed of sound, the mechanical resonance occurs at $\omega_m/2\pi = 3.68$~GHz. In order to minimize mechanical damping in the structure, an external acoustic radiation shield is added in the periphery of the nanobeam (Fig.~\ref{fig:OM_device}d-e).  This acoustic shield consists of a two-dimensional ``cross'' pattern, which has been shown both theoretically and experimentally to yield a substantial phononic bandgap in the GHz frequency band~\cite{Alegre2011}.

\begin{figure}[h]
\begin{center}
\includegraphics[width=0.6\columnwidth]{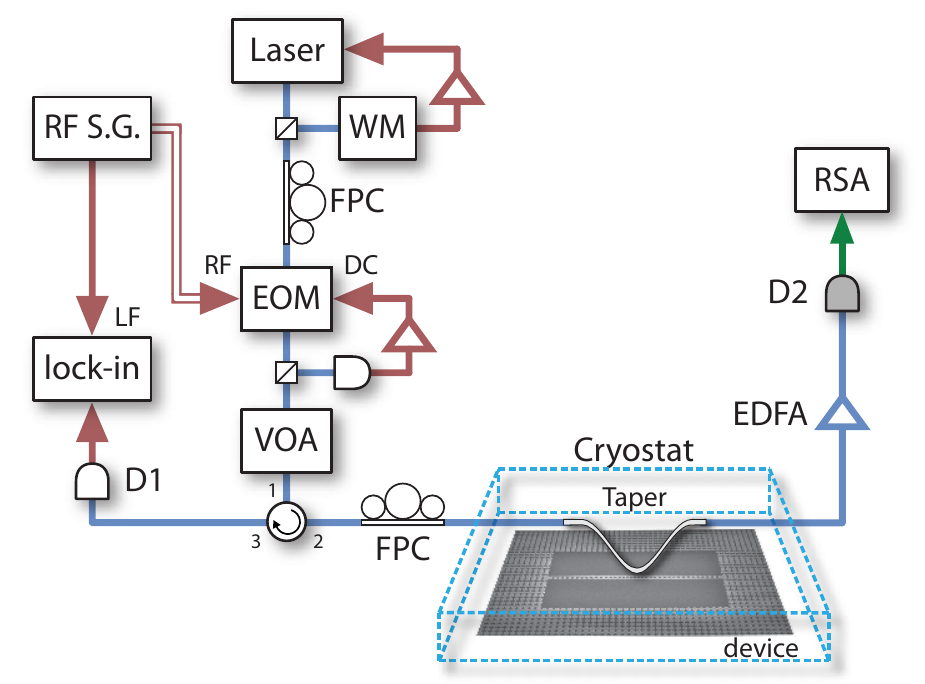}
\caption{\textbf{Experimental setup}. 
A single tunable 1550~nm diode laser is used as the cooling and mechanical transduction beam sent into the nanobeam optomechanical resonator cavity held in a continuous flow Helium cryostat. A wavemeter (WM) is used to track and lock the laser frequency, while a variable optical attenuator (VOA) is used to set the laser power. The transmitted signal is amplified by an erbium doped fiber amplifier (EDFA), and detected on a high-speed photodetector (D2) connected to a real-time spectrum analyzer (RSA), where the mechanical noise power spectrum is measured. A slowly modulated probe signal used for optical spectroscopy and calibration is generated from the cooling laser beam via an amplitude electro-optic modulator (EOM) driven by a microwave source (RFSG).  The reflected component of this signal is separated from the input via an optical circulator, sent to a photodetector (D1), and then demodulated on a lock-in amplifier.  Paddle-wheel fiber polarization controllers (FPCs) are used to set the laser polarization at the input to the EOM and the input to the optomechanical cavity. For more detail see Appendix~\ref{AppD}.}
\label{fig:setup}
\end{center}
\end{figure}

A fiber taper nanoprobe, formed from standard single-mode optical fiber, is used to optically couple to the silicon nanoscale resonators. As shown in Fig.~\ref{fig:setup}, a tunable laser (New Focus Velocity swept laser; $200$~kHz linewidth) is used to optically cool and transduce the mechanical motion of the nanomechanical oscillator.  Placing the optomechanical devices into a continuous-flow helium cryostat provides a modicum of pre-cooling down to $T_b\approx20$~K, reducing the bath occupancy of the $3.68$~GHz mechanical mode to $n_{b} \approx 100$.  At this temperature the mechanical $Q$-factor increases up to a measured value of $Q_m\approx 10^5$, corresponding to an intrinsic mechanical damping rate of $\gamma_i/2\pi=35$~kHz.  The optical $Q$-factor is measured to be $Q_{o}=4\times10^5$, corresponding to an optical linewidth of $\kappa/2\pi=500$~MHz, slightly reduced from its room temperature value.

In the resolved sideband limit in which $\omega_m/\kappa > 1$, driving the system with a laser (frequency $\omega_l$) tuned to the red side of the optical cavity (detuning $\Delta\equiv\omega_o-\omega_l = \omega_m$), creates an optically-induced damping, $\gamma_\text{OM}$, of the mechanical resonance~\cite{Wilson-Rae2007,Marquardt2007}. In the weak-coupling regime ($\gamma_\text{OM} \ll \kappa$) and for a detuning $\Delta = \omega_m$, the optical back-action damping is given by $\gamma_\text{OM}=4g^2n_c/\kappa$, where $n_c$ is the average number of drive photons stored in the cavity and $g$ is the optomechanical coupling rate between the mechanical and optical modes.  This coupling rate, $g$, is quantified as the shift in the optical resonance for an amplitude of motion equal to the zero-point amplitude ($x_\text{zpf}=(\hbar/2m\omega_m)^{1/2}$; $m$ the motional mass of the localized acoustic mode, $\hbar$ Planck's constant divided by $2\pi$).  The optomechanical damping, a result of the preferential scattering of drive laser photons into the upper-frequency sideband, also cools the mechanical mode.  For a quantum-limited drive laser, the phonon occupancy of the mechanical oscillator can be reduced from $n_b=k_BT_b/\hbar\omega_m \gg 1$, to a value $\bar{n}=n_b/(1+C) + n_{\text{min}}$, where $C\equiv \gamma_{\text{OM}}/\gamma_{i}$ is the cooperativity. The residual scattering of drive laser photons into the lower-frequency sideband limits the cooled phonon occupancy to $n_{\text{min}} = \left(\kappa/4\omega_m\right)^2$, determined by the level of sideband resolution~\cite{Wilson-Rae2007,Marquardt2007}.
  
\begin{figure}[h]
\begin{center}
\includegraphics[width=0.7\columnwidth]{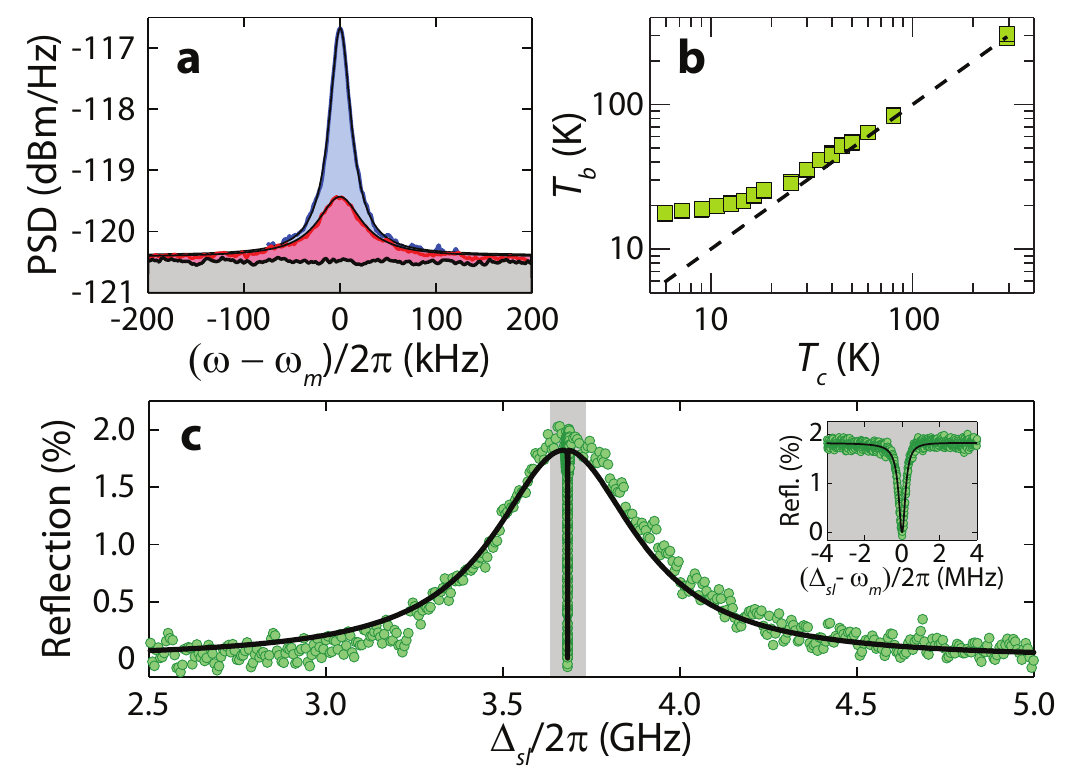}
\caption{\textbf{Mechanical and optical response}. \textbf{a}, Typical measured mechanical noise spectra around the resonance frequency of the breathing mode for low laser drive power ($n_{c} = 1.4$).  The blue (red) curve corresponds to the measured spectrum with laser drive blue (red) detuned by a mechanical frequency from the optical cavity resonance.  The black trace corresponds to the measured noise floor (dominated by EDFA noise) with the drive laser detuned far from cavity resonance. \textbf{b}, Plot of the measured ($\square$) mechanical mode bath temperature  ($T_b$) versus cryostat sample mount temperature ($T_c$). The dashed line indicates the curve corresponding to perfect following of the mode temperature with the cryostat temperature ($T_b=T_c$). \textbf{c}, Typical reflection spectrum of the cavity while driven by the cooling laser at $\Delta = \omega_m$ as measured by a weaker probe beam at two-photon detuning $\Delta_{sl}$. The signature reflection dip on-resonance with the bare cavity mode, highlighted in the inset, is indicative of electromagnetically-induced transparency (EIT) caused by the coupling of the optical and mechanical degrees of freedom by the cooling laser beam. }
\label{fig:spectra}
\end{center}
\end{figure}

The drive laser, in addition to providing mechanical damping and cooling, can be used to measure the mechanical and optical properties of the system through a series of calibrated measurements.  In a first set of measurements the noise power spectral density (PSD) of the drive laser transmitted through the optomechanical cavity is used to perform spectroscopy of the mechanical mode.  As shown in Appendix~\ref{AppA}, the noise power spectral density of the photocurrent generated by the transmitted field of the drive laser with red-sideband detuning ($\Delta=\omega_m$) yields a Lorentzian component of the single-sided PSD proportional to ${S}_{b}(\omega)={\bar{n}\gamma}/{((\omega-\omega_{m})^2 + (\gamma/2)^2)},$ where $\gamma=\gamma_i + \gamma_\text{OM}=\gamma_i(1+C)$ is the total mechanical damping rate. For a blue laser detuning of $\Delta=-\omega_m$, the optically-induced damping is negative ($\gamma_\text{OM}=-4g^2n_c/\kappa$) and the photocurrent noise PSD is proportional to ${S}_{b^\dag}(\omega)={(\bar{n}+1)\gamma}/\left({(\omega-\omega_{m})^2 + (\gamma/2)^2}\right)$. Typical measured noise power spectra under low power laser drive ($n_c = 1.4$, $C = 0.27$), for both red and blue detuning, are shown in Fig.~\ref{fig:spectra}a. Even at these small drive powers the effects of back-action are clearly evident on the measured spectra, with the red-detuned drive broadening the mechanical line and the blue-detuned drive narrowing the line. The noise floor in Fig.~\ref{fig:spectra}a (shaded in gray) corresponds to the noise generated by the erbium doped fiber amplifier (EDFA) used to pre-amplify the transmitted drive laser signal prior to photodetection, and is many orders of magnitude above the electronic noise of the photoreceiver and real-time spectrum analyzer.

\begin{figure*}[ht!]
\begin{center}
\includegraphics[width=\columnwidth]{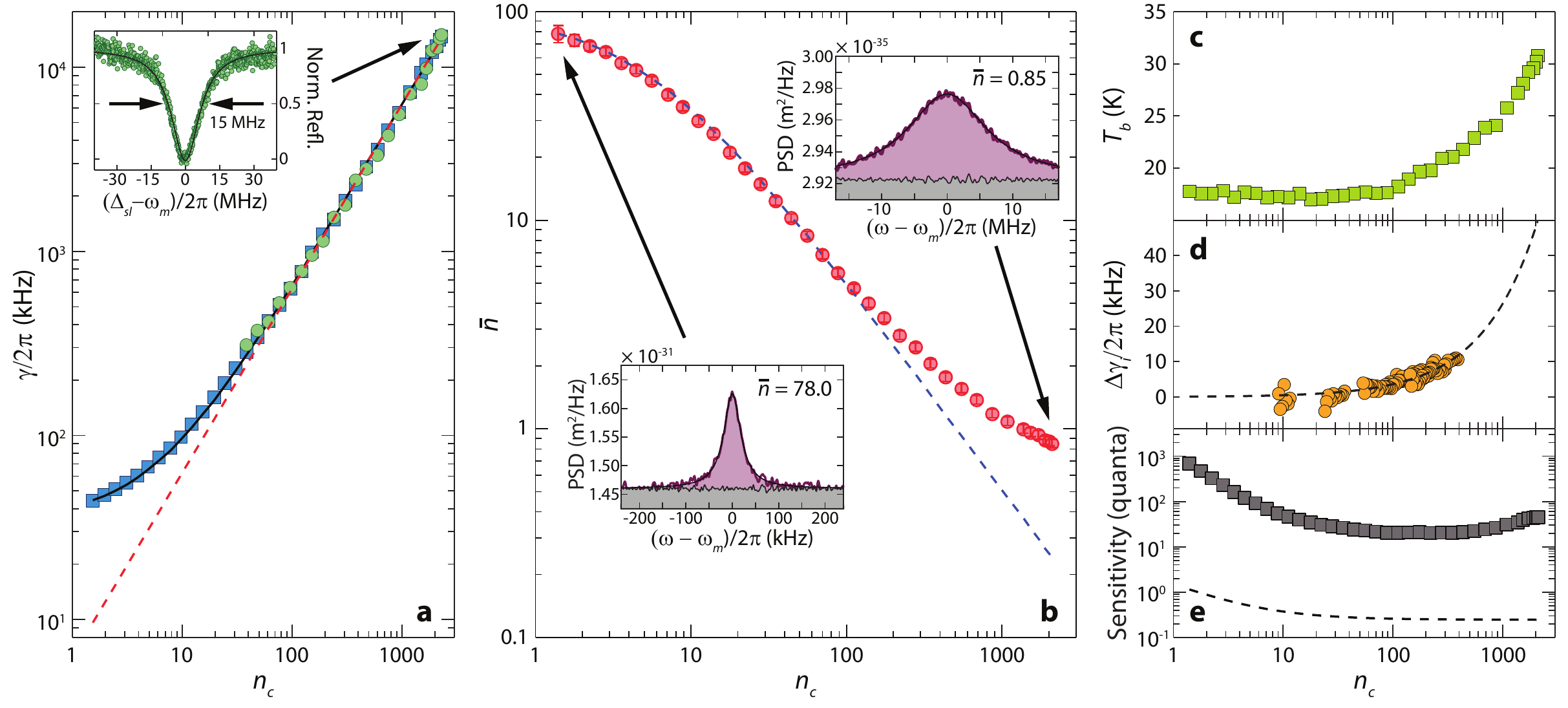}
\caption{\textbf{Optical cooling results}.  \textbf{a}, Measured mechanical mode linewidth ($\square$), EIT transparency bandwidth ($\circ$), and predicted optomechanical damping rate estimated using the zero-point optomechanical coupling rate $g/2\pi=910$~kHz (red dashed line). The inset shows the measured EIT transparency window at the highest cooling drive power. \textbf{b}, Measured ($\circ$) average phonon number, $\bar{n}$, in the breathing mechanical mode at $\omega_m/2\pi=3.68$~GHz versus cooling laser drive power (in units of intracavity photons, $n_c$), as deduced from the calibrated area under the Lorentzian lineshape of the  mechanical noise power spectrum, ${S}_{b}$. The left and right inset spectra correspond to the measured noise power spectrum in units of m$^2$/Hz at low and high laser cooling power, respectively.   The dashed blue line indicates the estimated mode phonon number from the measured optical damping alone. Error bars indicate computed standard deviations as outlined in the SI. \textbf{c}, Estimated bath temperature, $T_b$, versus cooling laser intracavity photon number, $n_c$. \textbf{d}, Measured change in the intrinsic mechanical damping rate versus $n_c$ ($\circ$). A polynomial fit to the mechanical damping dependence on $n_c$ is shown as a dashed line. For more detail see the SI. \textbf{e},  The measured ($\square$) background noise PSD versus laser drive power ($n_c$), in units of phonon quanta.  Shown as a black dashed line is the theoretical shot-noise limited noise PSD for an ideal single-sided cavity, a unit quantum efficiency detector, and no optical loss in the transmitted optical field. }
\label{fig:cooling}
\end{center}
\end{figure*}

Calibration of the EDFA gain, along with the photoreceiver and real-time spectrum analyzer photodetection gain, allows one to convert the measured area under the photocurrent noise PSD into a mechanical mode phonon occupancy. As described in detail in the Appendices, these calibrations, along with measurements of low drive power ($C \ll 1$) rf-spectra of both $\Delta=\pm\omega_m$ detunings, are performed to provide an accurate, local thermometry of the optomechanical cavity. An example of this form of calibrated mode thermometry is shown in Fig.~\ref{fig:spectra}b, where we plot the optically measured mechanical mode bath temperature ($T_b$) versus the cryostat sample mount temperature ($T_c$; independently measured using a Si diode thermometer attached to the copper sample mount). As one can see from this plot, the optical mode thermometry accurately predicts the absolute temperature of the sample for $T_c > 50$~K; below this value the mode temperature deviates from $T_c$ and saturates to a value of $T_b=17.6\pm 0.8$~K due to black-body heating of the device through the imaging aperture in the radiation shield of our cryostat.

In a second set of measurements, the mechanical damping, $\gamma$, and the cavity-laser detuning, $\Delta$, can be measured by optical spectroscopy of the driven cavity. By sweeping a second probe beam of frequency $\omega_s$ over the cavity, with the cooling beam tuned to  $\Delta=\omega_m$, spectra exhibiting electromagnetically-induced transparency (EIT)~\cite{Weis2010,Teufel2011a,Safavi-Naeini2011} are measured, as shown in Fig.~\ref{fig:spectra}c.  Due to the high single-photon cooperativity in the system, an intracavity population of only $n_c \approx 5$ switches the system from reflecting to transmitting for the probe beam.  The corresponding dip at the center of the optical cavity resonance occurs at a two-photon detuning $\Delta_{sl} \equiv \omega_s - \omega_l = \omega_m$ and has a bandwidth equal to the mechanical damping rate, $\gamma_i(1+C)$.  Figure~\ref{fig:cooling}a shows a plot of the measured mechanical linewidth versus intracavity photon number, displaying good correspondence between both mechanical and optical spectroscopy techniques.  From a fit to the measured mechanical damping rate versus $n_c$ (dashed red line in Fig.~\ref{fig:cooling}a), the zero-point-motion optomechanical coupling rate is estimated to be $g/2\pi=910$~kHz, placing the system well within the weak-coupling regime for all measured drive powers.   

In Fig.~\ref{fig:cooling}b we plot the calibrated Lorentzian noise PSD area, in units of phonon occupancy, versus red-detuned ($\Delta=\omega_m$) drive laser power.  Due to the low effective temperature of the laser drive, the mechanical mode is not only damped but also cooled substantially. The minimum measured mode occupancy for the highest drive power of $n_c\approx 2000$ is $\bar{n}=0.85\pm0.04$, putting the mechanical oscillator in a thermal state with ground state occupancy probability greater than 50\%.  The dashed blue line in Fig.~\ref{fig:cooling}b represents the ideal back-action cooled phonon occupancy estimated using both the measured mechanical damping rate in Fig.~\ref{fig:cooling}a and the low drive power intrinsic mechanical damping rate.  Deviation of the measured phonon occupancy from the ideal cooling model is seen to occur at the highest drive powers, and as detailed in Appendices~\ref{AppI}-\ref{AppK}, is due to both an increase in the bath temperature due to optical absorption (Fig.~\ref{fig:cooling}c) and an increase in the intrinsic mechanical damping rate (Fig.~\ref{fig:cooling}d) induced by the generation of free-carriers through optical absorption.  In order to evaluate the efficiency of the optical transduction of the mechanical motion, we also plot in Fig.~\ref{fig:cooling}e the measured background noise PSD (or imprecision level) alongside that for an ideal cavity transducer with shot-noise limited detection.  The minimum measurement imprecision corresponds to $n_{\text{imp}} \approx 20$ in units of phonon quanta, $40$\% of which is due to light lost inside the optical cavity (and thus not detected).  The remaining $60$\% of the imprecision stems from optical loss in the fiber taper (approximately 2~dB) and added noise due to the EDFA pre-amplification. 

Looking ahead, the combination of strong optical back-action cooling and efficient mechanical motion transduction realized in the chip-scale optomechanical cavities of this work, represents a first step towards optical quantum control of nanomechanical objects~\cite{aspelmeyer2010b}.  For example, optomechanical entanglement between light and mechanics~\cite{Vitali2007} or quantum state transfer between single optical photons and mechanical phonons~\cite{Akram2010,Safavi-Naeini2011a} may be envisioned, enabling mechanical systems to function as either quantum transducers~\cite{Stannigel2010} or quantum memory elements~\cite{Chang2011}.  The efficacy of such quantum protocols and devices, and the ability to measure quantum dynamics, relies on the thermal decoherence time of the mechanical system, given by $\tau \equiv k_BT_b/\hbar Q_m$.  For the measured devices in this work at $T_b\approx 20$~K, the decoherence time corresponds to $N_{\text{osc}}\equiv\tau\omega_m/2\pi \approx 200$ periods of coherent oscillation of the mechanical resonator.  Back-action cooling\cite{Marquardt2007} from a given bath temperature to the quantum ground state requires $\hbar\kappa \gtrsim k_B T_b/Q_m$.  Given the properties of the devices in this work, ground state cooling from room temperature with measurable quantum dynamics ($N_{\text{osc}}\approx 10$) seems feasible.  This opens up the possibility for future experimentation with, and utilization of, quantum nanomechanical objects in a room temperature environment. 

\section*{Acknowledgements}

This work was supported by the DARPA/MTO ORCHID program through a grant from AFOSR, the European Commission (MINOS, QUESSENCE), European Research Council (ERC QOM), the Austrian Science Fund (CoQuS, FOQUS, START),  and the Kavli Nanoscience Institute at Caltech. JC and ASN gratefully acknowledge support from NSERC.   


\appendix

\section{Classical Derivation of Transduced Signal}
\label{AppA}
 We begin by modeling the optomechanical system with the Hamiltonian
\begin{equation}
\hat H = \hbar\Delta \hat a^\dagger\hat a + \hbar\omega_m \hat b^\dagger\hat b + \hbar g(\hat b^\dagger + \hat b)\hat a^\dagger \hat a + i\hbar\sqrt{\frac{\kappa_e}{2}}\alpha_\text{in,0}(\hat a - \hat a^\dagger),
\end{equation}
where $\Delta = \omega_o - \omegap$, with laser frequency $\omegap$, optical mode frequency $\omega_o$ and mechanical mode frequency $\omega_m$. Here $\hat{a}$ ($\hat a^\dagger$) and $\hat b$ ($\hat b^\dagger$) are respectively the annihilation (creation) operators of photon and phonon resonator quanta, $g$ is the optomechanical coupling rate corresponding physically to the shift in the optical mode frequency due to the zero-point fluctuations ($x_\text{zpf}=\sqrt{\hbar/2m\omega_m}$, $m$ motional mass) of the phonon mode. By making the substitutions
\begin{equation}
\hat a \rightarrow \alpha = \sum_q \alpha_q e^{-i q\omega_m t},\quad \hat b \rightarrow \beta_0 e^{-i \omega_m t}
\end{equation}
we can treat the system classically by representing the photon amplitudes as a Fourier decomposition of sidebands. Notice that the infinite summation over each sideband order $q$, can be relaxed to a few orders in the sideband resolved regime ($\kappa\ll\omega_m$). The phonon amplitude, $\beta_0$, is the classical mechanical excitation amplitude. For an oscillator undergoing thermal Brownian motion, $\beta_0$, is a stochastic process. We assert the stochastic nature of the variable, at the end of the derivation where the power spectral density is calculated.  The equation of motion for the slowly varying component is then
\begin{equation}
-i\omega_m \sum_{q} q \alpha_q e^{-iq \omega_m t} = -\left(i\Delta + \frac\kappa2\right)\sum_q \alpha_q e^{-iq\omega_m t} - i g \beta_0 \sum_q\alpha_q \left(e^{-i(q+1)\omega_m t}+e^{-i(q-1)\omega_m t} \right) - \sqrt{\frac{\kappa_e}{2}}\alpha_\text{in,0},
\end{equation}
where we introduce the cavity (optical) energy loss rate, $\kappa$, and the cavity coupling rate, $\kappa_e$. This can be written as a system of equations $\mathbf{M}\cdot\vec{\alpha} = \mathbf{a_{in}}$ where
\begin{align}
M_{pq} &= \left(i(\Delta-p\omega_m) + \frac\kappa2\right)\delta_{pq} + ig\beta_0(\delta_{p,q+1}+\delta_{p,q-1}),\\
\mathrm{a_{in}}_{,p} & = -\sqrt{\frac{\kappa_e}{2}}\alpha_\text{in,0} \delta_{p0}.
\end{align}
By truncating and inverting the coupling matrix $\mathbf{M}$ one can determine each one of the sidebands amplitude as $\alpha_q=(M^{-1})_{qp}~\mathrm{a}_{\text{in},p}$ and therefore determine the steady state power leaving the cavity to be
\begin{equation}
\alpha_\text{out} = \alpha_\text{in,0} + \sqrt{\frac{\kappa_e}{2}}\alpha
\end{equation}
where we assumed that the input pump is not depleted by the cavity, which is the frame of interest of this work. In this case the total power measured at the photodetector will be proportional to
\begin{align}
|\alpha_\text{out}|^2 &= \left| \alpha_\text{in,0}+ \sqrt{\frac{\kappa_e}{2}}\alpha \right|^2\\
&= |\alpha_\text{in,0}|^2 + \frac{\kappa_e}{2}\sum_q\sum_p \alpha_q\alpha_p^*e^{-i(q-p)\omega_mt} + 2\mbox{Re}\left\{ \alpha_\text{in,0}\sqrt{\frac{\kappa_e}{2}}\sum_q \alpha_q e^{iq\omega_mt} \right\}
\end{align}

\subsection{Resolved Sideband Limit}
The equations presented in the previous section are exact and therefore can be solved for any case. However our interests lie in the so-called resolved sideband limit, $\omega_m>\kappa/2$, where further simplification can be done. Specifically for our system, $\omega_m/2\pi=3.68$~GHz, $\kappa/2\pi=500$~MHz putting us well within this limiting case.

Additionally, in the cavities studied, the optomechanical phase-modulation factor (proportional to $g x_\text{zpf} / \omega_m$), is much less than $10^{-3}$. As such, only the $\omega=\omegap\pm\omega_m$ sidebands ($q = \pm 1$) are significant and $\alpha_0\gg\alpha_\pm$. Truncating the matrix equations appropriately, we find
\begin{align}\label{eqn:alpha_0}
\alpha_0 &= \frac{-\sqrt{\kappa_e/2}\alpha_\text{in,0}}{i\Delta+\kappa/2}\\
\alpha_\pm &= \frac{-i g\beta_0\alpha_0}{i(\Delta\mp\omega_m)+\kappa/2}
\end{align}
where $\alpha_\text{in,0}=\sqrt{N_\text{in}}$, $N_\text{in}=P_\text{in}/\hbar\omega_0$ and $P_\text{in}$ the input power at the cavity.

In the sideband resolved limit, we have $|\alpha_\text{in,0}|>|\sqrt{\kappa_e/2}\alpha_0|>|\sqrt{\kappa_e/2}\alpha_\pm|$.  Therefore the photodetector signal is predominantly composed of the mixing between sidebands with the input pump beam and terms containing $|\alpha_0|^2$, and can be written as~\cite{Eichenfield2009b}:
\begin{align}\nonumber
|\alpha_\text{out}|^2 &=  |\alpha_\text{in,0}|^2+ \sqrt{\frac{\kappa_e}{2}}\alpha_\text{in,0}(\alpha_0+\alpha_0^\ast) + \frac{\kappa_e}{2}|\alpha_0|^2 + ... \\\nonumber
&\quad~\sqrt{\frac{\kappa_e}{2}}\alpha_\text{in,0}(\alpha_-e^{-i\omega_mt}+\alpha_-^\ast e^{i\omega_mt})+...\\
&\quad~\sqrt{\frac{\kappa_e}{2}}\alpha_\text{in,0}(\alpha_+e^{i\omega_mt}+\alpha_+^\ast e^{-i\omega_mt})+ \mathcal{O}(|\alpha_0||\alpha_\pm|) \\\nonumber
&\approx|\alpha_\text{in,0}|^2\left|1 - \frac{\kappa_e/2}{i\Delta+\kappa/2}\right|^2 + ...\\\nonumber
&\quad~\cos(\omega_mt)\Big[|A_+|\cos(\varphi_+)+|A_-|\cos(\varphi_-)\Big] + ...\\\label{eqn:sout}
&\quad~\sin(\omega_mt)\Big[|A_+|\sin(\varphi_+)-|A_-|\sin(\varphi_-)\Big]
\end{align}
where $A_{\pm} \equiv 2\sqrt{\kappa_e/2}~\alpha_\text{in,0}\alpha_\pm = |A_{\pm}|\exp{(-i\varphi_\pm)}$. We can easily recognize the first term in Equation~(\ref{eqn:sout}) as the DC cavity transmission spectra. The remaining two terms compose the total power at the mechanical frequency $P_\text{SB}(\omega_m)=\hbar\omega_0\sqrt{A_\text{cos}^2+A_\text{sin}^2}$, where $A_\text{cos}=|A_+|\cos(\varphi_+)+|A_-|\cos(\varphi_-)$ and $A_\text{sin}=|A_+|\sin(\varphi_+)-|A_-|\sin(\varphi_-)$.

Given a mechanical system which is oscillating coherently at a frequency $\omega_m$ ($\beta_0$ is simply a complex number) the single sided spectral density of the power at the detector, as a function of the laser detuning $\Delta$, and frequency $\omega$, will be given by
\begin{eqnarray}\label{eqn:PSB_out}
S_\text{PP}(\omega, \Delta)&=&\hbar^2 \omega_0^2 \kappa_e^2 \big|\alpha_\text{in,0}\big|^4\times\Big|\frac{ig\beta_0}{(i\Delta+\kappa/2)(i(\Delta-\omega_m)+\kappa/2)}\Big|^2\times\delta(\omega - \omega_m)\\\nonumber
&=&\hbar^2\omega_0^2\frac{g^2|\beta_0|^2\kappa_e^2\big|\alpha_\text{in,0}\big|^4}{(\Delta^2+(\kappa/2)^2)((\Delta-\omega_m)^2+(\kappa/2)^2)}\times \delta(\omega - \omega_m).
\end{eqnarray}

For mechanical systems undergoing random oscillations, the important quantity is the power spectral density of the detected signal. Since this is calculated from the autocorrelation functions, it will contain products of the form $\beta_0^\ast(t) \beta_0(t')$. Classically, these averages may be calculated from the Boltzmann distribution, and can be replaced with $\bar{n}_T=k_B T_b /\hbar\omega_m$, with $k_B$ the Boltzmann constant, and $T_b$ the bath temperature.  Additionally, since the measured sideband is blue of the pump frequency ($\omegap+\omega_m$), from the quantum theory~\cite{Wilson-Rae2007,Marquardt2007}, and the derivation below, the proper ordering to be used is the normal one ($b^\dagger b$), and thus the expectation values may be replaced with $\bar{n}$, the number of phonons occupying the mechanical mode. As such, we can effectively use the derivations shown above, in both the classical and quantum cases, substituting $\beta_0$ by $\sqrt{\bar{n}}$, and replacing the delta functions $\delta(\omega - \omega_m)$ with unit-area Lorentzian functions. A fully quantum mechanical derivation of this result is shown below.

\subsection{Quantum Mechanical Derivation of Observed Spectra}

In this section we use the following conventions for Fourier transforms and spectral densities. Given an operator $A$, we take
\begin{eqnarray}
\op{A}{}(t) &=& \frac{1}{\sqrt{2\pi}} \int_{-\infty}^{\infty} d\omega~ e^{-i\omega t}\op{A}{}(\omega), \nonumber\\
\op{A}{}(\omega) &=& \frac{1}{\sqrt{2\pi}} \int_{-\infty}^{\infty} dt~ e^{i\omega t}\op{A}{}(t),\nonumber\\
S_{AA}(\omega) &=& \int_{-\infty}^{\infty} d\tau~ e^{i\omega \tau} \avg{\opdagger{A}{}(t+\tau)\op{A}{}(t)}. \nonumber
\end{eqnarray}
Additionally we define the symmetrized spectral density as $\bar{S}_{AA}(\omega) = \frac{1}{2} (S_{AA}(\omega) + S_{AA}(-\omega))$, and one-sided spectral densities which are those measured by the spectrum analyzer as $\bar{S}_{A}(\omega) = 2\bar{S}_{AA}(\omega)$. Starting from the quantum-optical Langevin equations for the mechanical ($\op{b}{}$) and optical ($\op{a}{}$) annihilation operators,
\bea
\dot{\op{b}{}}(t) &=& -\left(i\omega_m + \frac{\gamma_i}{2}\right) \op{b}{} - ig \opdagger{a}{} \op{a}{} - \sqrt{\gamma_i}\op{b}{\mathrm{in}}\qquad\mathrm{and}\\
\dot{\op{a}{}}(t) &=& -\left(i\Delta + \frac{\kappa}{2}\right) \op{a}{} - ig \op{a}{}(\opdagger{b}{}+\op{b}{}) -\sqrt{\kappa_e/2} \op{a}{\mathrm{in}}(t) - \sqrt{\kappa^\prime}\op{a}{\mathrm{in},i}(t),
\eea
we linearize the equations about a large optical field intensity by displacing $\op{a}{} \rightarrow \alpha_0 + \op{a}{}$. Then in Fourier domain the fluctuations are then given by
\bea
\op{b}{}(\omega) &=&  \frac{-\sqrt{\gamma_i}\op{b}{\mathrm{in}}(\omega)}{i(\omega_m-\omega) + \gamma_i/2} -  \frac{i G(\op{a}{}(\omega) + \opdagger{a}{}(\omega))}{i(\omega_m-\omega) + \gamma_i/2},\label{eqn:mech_fluct}\\
\op{a}{}(\omega) &=&  \frac{-\sqrt{\kappa_e/2} \op{a}{\mathrm{in}}(\omega) - \sqrt{\kappa^\prime}\op{a}{\mathrm{in},i} - iG(\op{b}{}(\omega) + \opdagger{b}{}(\omega))  }{i(\Delta-\omega) + \kappa/2}\label{eqn:opt_fluct},
\eea
where $\kappa^\prime = \kappa - \kappa_e/2$ denotes all the optical loss channels which go undetected (i.e. information is lost) and $G = g\alpha_0$. 
For an ideal measurement, $\kappa_e/2 = \kappa$, and $\kappa^\prime = 0$, so the intrinsic vacuum fluctuations ($\op{a}{\mathrm{in},i}$) never enter the optical cavity. For a double sided coupling scheme, such as the one with a fiber taper, $\kappa = \kappa_i + \kappa_e$, and so $\kappa^\prime = \kappa_e/2$ at best, due to the back reflection from the cavity, which contains information about the mechanics which is lost.

We account for all the fluctuations (vacuum and thermal) incident on the photodetector, and calculating the spectra of each term, we find the heterodyne detected signal.
Using Equations (\ref{eqn:mech_fluct}) and (\ref{eqn:opt_fluct}) we arrive at the operator for the mechanical fluctuations,
\bea
\op{b}{}(\omega) =  \frac{-\sqrt{\gamma_i}\op{b}{\mathrm{in}}(\omega)}{i(\omega_m-\omega) + \gamma/2}~~~~~~~~~~~~~~~~~~~~~~~~~~~~~~~\nonumber\\
+ \frac{iG}{i(\Delta -\omega) + \kappa/2} \frac{\sqrt{\kappa_e/2} \op{a}{\mathrm{in}}(\omega) + \sqrt{\kappa^\prime} \op{a}{\mathrm{in},1}(\omega)}{i(\omega_m-\omega) + \gamma/2}\nonumber\\
+ \frac{iG}{-i(\Delta + \omega) + \kappa/2} \frac{\sqrt{\kappa_e/2} \opdagger{a}{\mathrm{in}}(\omega) + \sqrt{\kappa^\prime} \opdagger{a}{\mathrm{in},i}(\omega)}{i(\omega_m-\omega) + \gamma/2}\label{eqn:b_inputs}
\eea
where $\omega_m$ is now the optical-spring shifted mechanical frequency, and $\gamma = \gamma_i + \gamma_\mathrm{OM}$, the optically damped mechanical loss-rate.

\subsubsection{Simplified Result under RWA ($\kappa^2/16\omega_m^2 \ll 1$) and Weak-Coupling ($G \ll \kappa$)}

Assuming that $\Delta = \omega_m$ and that we care mainly about the system response around $\omega_m$ (where $\op{b}{}(\omega)$ is peaked), the relation (\ref{eqn:b_inputs}) can be simplified, 
\bea
\op{b}{}(\omega) &=&  \frac{-\sqrt{\gamma_i}\op{b}{\mathrm{in}}(\omega)}{i(\omega_m-\omega) + \gamma/2} + \frac{2iG}{\kappa} \frac{\sqrt{\kappa_e/2} \op{a}{\mathrm{in}} + \sqrt{\kappa^\prime} \op{a}{\mathrm{in},i}(\omega)}{i(\omega_m-\omega) + \gamma/2} + \mathcal{O}\left( \frac{G}{2\omega_m} \right)
\eea
and we drop the term $\propto \frac{1}{2\omega_m}$ (RWA).

We find using the input-output boundary condition 
\bea
\op{a}{\mathrm{out}}(\omega) &=& \op{a}{\mathrm{in}}(\omega)  + \sqrt{\kappa_e/2} \op{a}{}(\omega) +  E_\mathrm{LO}\delta(\omega) \\
&=& \op{a}{\mathrm{in}}(\omega) \left( 1 - \frac{\kappa_e}{\kappa} 
+ \frac{4|G|^2}{\kappa} \frac{\kappa_e}{2\kappa}\frac{1}{i(\omega_m - \omega) + \gamma/2} \right) \nonumber \\
&~& +  \op{a}{\mathrm{in},i}(\omega) \left( -\sqrt{\frac{2\kappa^\prime\kappa_e}{\kappa^2}} + \frac{4|G|^2}{\kappa} \sqrt{\frac{\kappa^\prime \kappa_e}{2\kappa^2}}\frac{1}{i(\omega_m - \omega) + \gamma/2}   \right)\nonumber\\
&~& +  \op{b}{\mathrm{in}}(\omega) \left( i G \sqrt{\frac{2\gamma_i \kappa_e}{\kappa^2}}\frac{1}{i(\omega_m - \omega) + \gamma/2}  \right)+ E_\mathrm{LO}\delta(\omega) \nonumber \\
&=&  s_{11}(\omega) \op{a}{\mathrm{in}}(\omega) + n_{\mathrm{opt}}(\omega) \op{a}{\mathrm{in},i}(\omega) + s_{12}(\omega) \op{b}{\mathrm{in}}(\omega) +  E_\mathrm{LO} \delta(\omega) 
\eea
with the scattering matrix elements above defined as in Ref. \cite{Safavi-Naeini2011a}. For the case where the mechanical bath is at zero temperature, the spectral density will be given simply by $|s_{11}(\omega)|^2 + | n_{\mathrm{opt}}(\omega)|^2 + |s_{12}(\omega)|^2  = 1$, as a result of all input fluctuations being uncorrelated, and therefore no feature is present at the mechanical frequency.

For the case of the $n_b > 0$, we find the autocorrelation of the detected normalized photocurrent $\op{I}{}(t) =\op{a}{\mathrm{out}}(t) + \opdagger{a}{\mathrm{out}}(t)$ to be
\bea
S_{II}(\omega) &=& |s_{11}(\omega)|^2 + | n_{\mathrm{opt}}(\omega)|^2 + |s_{12}(\omega)|^2 (n_b+1) +  |s_{12}(-\omega)|^2 n_b\nonumber \\
&=& 1 + n_b(|s_{12}(\omega)|^2  + |s_{12}(-\omega)|^2 ) \nonumber \\
&=& 1 + \frac{\kappa_e}{2\kappa} \frac{4|G|^2}{\kappa} \left( \frac{\gamma_in_b/\gamma}{(\omega_m - \omega)^2 + (\gamma/2)^2} +  \frac{\gamma_in_b/\gamma}{(\omega_m + \omega)^2 + (\gamma/2)^2} \right)\nonumber\\
&=& 1 + \frac{\kappa_e}{2\kappa} \frac{8|G|^2}{\kappa}  \bar{S}_{bb}(\omega) \label{eqn:SII}
\eea
where for the last step we've used the fact that in the highly sideband-resolved regime, $\bar{n} = \gamma_in_b/\gamma$.

\subsection{Optomechanical Damping}
For the case where $\Delta = \omega_m$ (pumping on the red side of the cavity), the steady-state phonon amplitude can be written for a sideband resolved system far from strong coupling as~\cite{Marquardt2007,Wilson-Rae2007}
\be\label{eqn:n_cool}
\bar{n}= \frac{\gamma_i}{\gamma_\text{OM} + \gamma_i}n_b,
\ee
with $n_b$ the equilibrium mechanical mode occupation number determined by the mechanical bath temperature, $\gamma_i$ the mechanical coupling rate to the bath and
\be\label{eqn:gamma_OM}
\gamma_\text{OM}= \frac{4g^2|\alpha_0|^2}{\kappa},
\ee
the resonant optomechanical damping rate.

\section{Fabrication}
\label{AppB}
The nano-beam cavities were fabricated using a Silicon-On-Insulator wafer from SOITEC ($\rho=4$-$20$~$\Omega\cdot$cm, device layer thickness $t=220$~nm, buried-oxide layer thickness $2$~$\mu$m). The cavity geometry is defined by electron beam lithography followed by inductively-coupled-plasma reactive ion etching (ICP-RIE) to transfer the pattern through the $220~\text{nm}$ silicon device layer. The cavities were then undercut using a $\text{HF:H}_2\text{O}$ solution to remove the buried oxide layer, and cleaned using a piranha/HF cycle~\cite{Borselli2006}. The dimensions and design of the nanobeam will be discussed in detail elsewhere.

\section{Device Parameters}
\label{AppC}
Under vacuum and cryogenic parameters, the optical resonance was found to have $Q_o = 4 \times 10^5$ (corresponding to $\kappa=500$~MHz), $\omega_o/2\pi = 195$~THz (corresponding to $\lambda_o=1537$~nm), and a resonant transmission contrast of $\Delta T=25$\%. The mechanical mode was found to have $Q_m = 1.06\times10^5$ (corresponding to $\gamma_i=35$~kHz) and $\omega_m/2\pi=3.68$~GHz. The optomechanical coupling rate was found to be $g/2\pi=910$~kHz.

\section{Experimental Setup}
\label{AppD}
\begin{figure*}[ht!]
\begin{center}
\includegraphics[width=14cm]{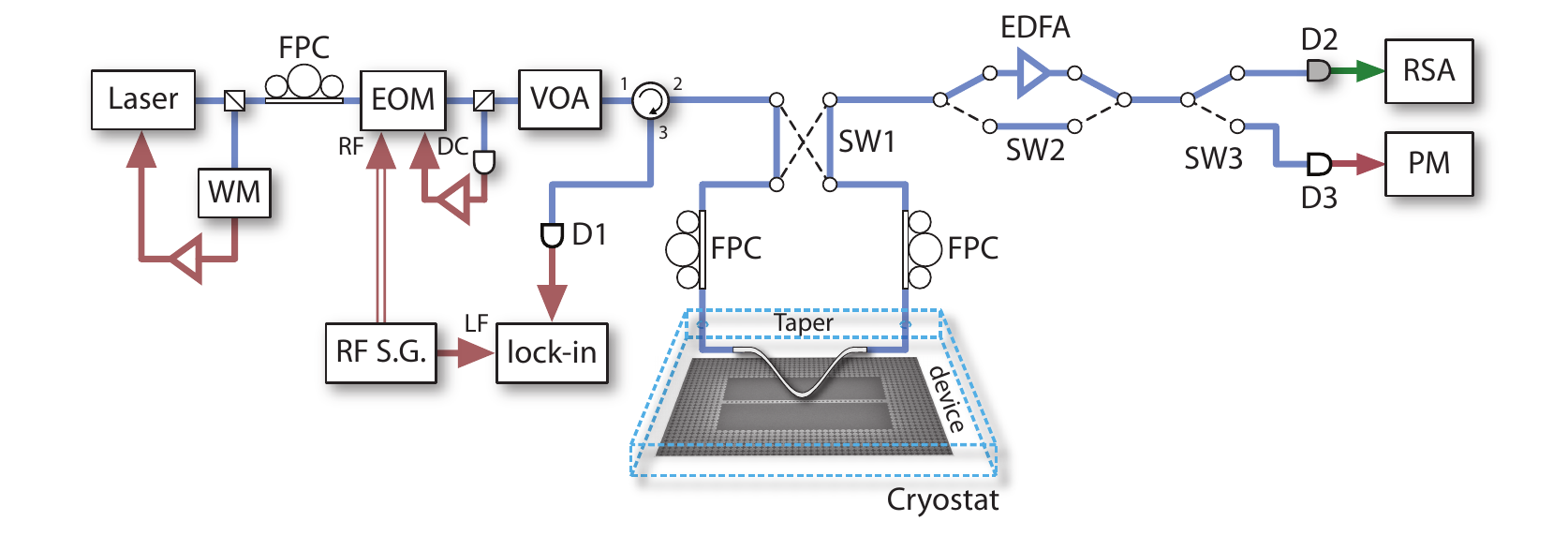}
\end{center}
\caption{\label{fig:fullsetup} Expanded experimental setup to include optical switches SW1, SW2 and SW3. The  blue lines indicate the optical path for the cooling measurement (the '0' position of each of the switches), while the dashed black lines indicate the alternative switched paths (the '1' position of each of the switches). A single tunable laser is used as the cooling laser and mechanical transduction laser.  A wavemeter (WM) is used to track and lock the laser's frequency to an absolute and relative value better than $100$~MHz and $5$~MHz, respectively. A calibrated (to better than $0.01$~dB) variable optical attenuator (VOA) is used to set the cooling laser power.  The transmitted component of the cooling laser beam that is sent into the optomechanical cavity is directed to an erbium doped fiber amplifier (EDFA), where the optical signal is pre-amplified before being detected on a high-speed photodetector (D2).  The measured photocurrent from D2 is sent to a real-time spectrum analyzer, where the mechanical noise power spectrum is measured.  A slowly modulated optical signal, on-resonance with the optical nanobeam cavity, is generated from the cooling laser beam via an amplitude electro-optic modulator (EOM) driven by a microwave source (RFSG).  The reflected component of the on-resonance laser signal injected into the cavity is separated from the input via an optical circulator, sent to a photodetector (D1), and then demodulated on a lock-in amplifier.  Paddle-wheel fiber polarization controllers (FPCs) are used to set the laser polarization at the input to the EOM and the input to the optomechanical cavity.}
\end{figure*}
The detailed experimental setup used to measure the cooling spectra and the electromagnetically-induced transparency (EIT) window of the optomechanical crystal is shown in Figure~(\ref{fig:fullsetup}). The setup is designed to measure both the EIT-like reflected signal and the transmission signal of the laser used to cool the mechanical system (though not simultaneously).

As a light source we use a fiber-coupled, tunable, near-infrared laser, (New Focus Velocity, model TLB-6328) spanning approximately $60$~nm centered around $1550$~nm, which has its intensity controlled by a variable optical attenuator (VOA). A small percentage ($10\%$) of the laser intensity is sent to wavemeter (WM, High Finesse, WS/6 High Precision) for passive frequency stabilization of the laser. To minimize polarization dependent losses on the electro-optical-modulator (EOM), a fiber polarization controller (FPC) is placed before it.

The EOM is driven by the microwave source (RF S.G., Agilent, model E8257D-520). The RF signal is composed of an amplitude modulated RF-signal carrier swept between $\Delta=1-8$~GHz modulated at the lock-in detection frequency, $\omega_\text{LI}$. As a result, the EOM modulation produces two probe sidebands at $\pm \Delta$, each with a small amplitude modulation at the lock-in frequency.

A small portion of the signal from the EOM output ($10\%$) is used as a DC control signal to keep the EOM level locked, compensating for any low frequency power drift during the experiment. The remaining laser light is passed through a circulator, a switch (SW1), and a fiber polarization controller (FPC). It is then coupled to a tapered and dimpled optical fiber (Taper) which has its position controlled with nanometer-scale precision.

Switches 2 and 3 (SW2 and SW3) determine the path that the light transmitted through the taper follows. In the normal configuration, the transmitted light is optically amplified by an erbium doped fiber amplifier (EDFA) and then detected by a high-speed photoreceiver (D2, New Focus, model 1554-B) which is connected to a real-time spectrum Analyzer (RSA, Tektronix RSA3408B). Detector 3 (D3) is used to measured the DC transmission response of the cavity. All the other configurations are used to calibrate the system as discussed in the calibration section.

Any reflected signal coming from the taper/device is detected by a high-gain photodetector (D1, New Focus, model 1811) and its signal is sent to a lock-in amplifier (L.I., SRS-830). The output from the in-phase and quadrature signals from the L.I. are recorded, producing the reflection scan shown in Figure~(3c) of the main text.

\section{Calibration for Mechanical Mode Thermometry}
\label{AppE}
To perform accurate mode thermometry, several optical switches were incorporated into the setup as shown in Figure~(\ref{fig:fullsetup}). A $2\times2$ switch (SW1) was positioned on the input/output ports of the fiber taper to control the direction of light through the taper, allowing the characterization of taper insertion loss asymmetry. Another $2\times2$ switch (SW2) was placed at the input/output ports of the EDFA, allowing the characterization of the optical gain, $G_\text{EDFA}$. Lastly, a $1\times2$ switch (SW3) was inserted before the RSA to switch the optical path between the RSA and a power meter (D3) that reads the total power, $P_\text{RSA}$, reaching the RSA, which allow us to monitor total insertion loss and provide a calibration for the electronic gain, $G_e$.

At the beginning of a measurement, the power into the taper from both SW1 paths is measured ($P_0$ and $P_1$ for the '0' and '1' position of the switch respectively). The insertion loss of the taper $L_\text{taper}$ is also measured. Finally, a pickoff power is recorded for each of the SW1 paths providing a correspondence between the measured taper input powers and the RSA powers, $P_\text{RSA,0}$ and $P_\text{RSA,1}$. From these calibration values, the insertion loss before ($L_0$) and after ($L_1$) the device (when the fiber taper is coupled) can be computed by assuming the bistability shift of the optical mode when sweeping the laser from blue to red is proportional to the dropped power. If we let primed values represent measurements made with the taper coupled to the device, and let $\Delta\lambda$ be the bistability shift, we have
\begin{equation}
\frac{P_{\text{in,0}}}{P_{\text{in,1}}} = \frac{P_0 L_0}{P_1 L_1} = \frac{\Delta\lambda_0}{\Delta\lambda_1}
\end{equation}
\begin{equation}
L_0L_1 = L_\text{taper}\frac{P_{\text{RSA},0}'}{P_{\text{RSA},0}}= L_\text{taper}\frac{P_{\text{RSA},1}'}{P_{\text{RSA},1}}
\end{equation}
from which $L_0$ and $L_1$ can be extracted. As a consequence, the intracavity power, $P_{\text{in}}$, can be accurately determined.

We directly measure $G_e$, by setting the attenuator to $0$~dB attenuation to measure $P_\text{RSA}$ and the corresponding DC bias voltage $V_\text{DC}$, and defining $G_e=V_\text{DC}/P_\text{RSA}$. The purpose of this is a technical one; the dynamic range of an optical power meter is much larger than that of a voltmeter. This allows us to accurately determine the DC bias voltage of the detector for any amplitude of optical signal through
\begin{equation}
V_\text{DC}' = V_\text{DC}\frac{P_\text{RSA}'}{P_\text{RSA}} = G_eP_\text{RSA}'.
\end{equation}
This value is critical because for photodetector signals of the form $V_{PD}(t) = V_\text{DC}'(1+\beta\sin\Omega t)$ where $\beta$ is the modulation depth and $\Omega$ is the modulation frequency, we have simply
\begin{equation}
\beta = \frac{\sqrt{2 P_\Omega R_L}}{V_\text{DC}'},  \label{eqn:beta1}
\end{equation}
where $P_\Omega$ is the integrated spectral power at $\Omega$, and we have assumed a detector load of $R_L$ and an RSA that reports $V_\text{RMS}$.

During the measurement, two calibration values are measured for each point in the power-dependent cooling run: $G_\text{EDFA}'$ and $P_\text{RSA}'$. The EDFA gain is measured by utilizing SW2 to insert and remove the EDFA from the optical train, while measuring a fixed tone at the mechanical frequency $\omega_m$ generated by the Electro-Optic Amplitude Modulator (EOM). The ratio of the integrated spectral power of the tones gives $G_\text{EDFA}'^2$. We also measure $P_\text{RSA}'$ without the EDFA in line so that when the EDFA is included, we have instead of Equation (\ref{eqn:beta1}),
\begin{equation}
\beta = \frac{\sqrt{2 P_\Omega R_L}}{G_\text{EDFA}'G_eP_\text{RSA}'}
\label{eqn:beta}
\end{equation}
to account for the additional optical gain.

Finally, by integrating the Lorentzian component of the power spectral density from Equation~(\ref{eqn:PSB_out}), we relate the detected signal on the spectrum analyzer to the calibrated values shown in Equation~(\ref{eqn:beta}). This allows us to make accurate determinations of the phonon number and mode temperature.

\section{EIT Measurements}
\label{AppF}
Here we will show how the amplitude modulation of the signal sideband $\tpd$ is used to measure the reflection ($|r(\omega)|^2$) of the signal reflected from the cavity. The output of the EOM can be written as:
\be
a_\text{out}(t) = a_\text{in}\Big[1+\beta \Big(1+m_\text{LI}\cos(\omega_\text{LI}t)\Big)\cos(\tpd t)\Big],
\ee
where the input field amplitude $a_\text{in}(t) = a_o\cos(\omegap t)$, $a_o=\sqrt{P_\text{in}/\hbar\omegap}$, $\beta$ is the EOM-modulation index, $\omega_\text{LI}$ and $m_\text{LI}$ are, respectively, the frequency and amplitude modulation index on the RF signal at $\tpd$. For the measurements shown in the main text $m_\text{LI}=1$. In this case one can write the field of the EOM output (cavity input) in the time domain as:
\begin{eqnarray}
 a_\textrm{out}(t) &=& a_o \Bigg[ \cos(\omegap t) + \frac{\beta}{2} \Big[ \cos((\omegap+\tpd)t) + \cos((\omegap-\tpd)t) \Big]  \nonumber \\
   & & + \frac{\beta}{4}       \Big[ \cos((\omegap + \tpd + \omega_\text{LI})t) + \cos((\omegap + \tpd - \omega_\text{LI}) t) \nonumber \\
   & & +                        \cos((\omegap - \tpd + \omega_\text{LI})t) + \cos((\omegap - \tpd - \omega_\text{LI}) t) \Big] \Bigg].
\end{eqnarray}
The reflected signal is filtered by the cavity dispersion and considering the case where the pump is on the red-side of the cavity ($\omegap<\omega_o$) the reflected field is:
\begin{eqnarray}
    a_{R}(t)=r(\omega_s)\frac{a_o\beta}{4}\Big[\cos((\omegap+\tpd) t)+\cos((\omegap+\tpd)t +(\omega_\text{LI}t -\varphi)) + \cos((\omegap+\tpd)t -(\omega_\text{LI}t -\varphi))\Big]
\end{eqnarray}
First we assume that $r(\omega_s)$ is roughly constant over a range of $\omega_\text{LI}$ which is true for $\omega_\text{LI}<(\gamma_i+\gamma_\text{OM})/2$. This implies that the smallest transparency window we could measure is on the order of the lock-in detection frequency. This limit on the transparency window size is reflected on Fig.~4c, where only transparency windows larger $200$~kHz are reported.

We can now write the time average detected power spectral density on the photoreceiver (D1 in Figure~(\ref{fig:fullsetup})) by taking the absolute square value of the reflected field and keeping only the terms with frequency smaller than the detector bandwidth. In this case:
\bea \nonumber
    P|_{\omega_s} &=& \frac{a_o^2\beta^2R_\text{PD}G_\text{PD}}{8R_L}|r(\omega_s)|^2\left[3+4\cos(\omega_\text{LI}t-\varphi)+\frac{1}{2}\cos(2\omega_\text{LI}t-2\varphi)+ \mathcal{O}(2\omegap)]\right].
\eea
where $R_\text{PD}=1$~A/W is the detector responsivity, $G_\text{PD}=40,000$~V/A is the detector gain and $R_L=50~\Omega$ is the load resistance.

This signal is then sent to the lock-in which can measure independently the in-phase ($X$) and quadrature ($Y$) power spectral densities at $\omega_\text{LI}$:
\bea \nonumber
    X|_{\omega_\text{LI}} &=& \frac{a_o^2\beta^2R_\text{PD}G_\text{PD}}{4R_L}|r(\omega_s)|^2\cos(\varphi)\\
    Y|_{\omega_\text{LI}} &=& \frac{a_o^2\beta^2R_\text{PD}G_\text{PD}}{4R_L}|r(\omega_s)|^2\sin(\varphi)
\eea
It is then easy to see the reflection amplitude and phase are given by
\bea \nonumber
    |r(\omega_s)|^2=\frac{4R_L}{a_o^2\beta^2R_\text{PD}G_\text{PD}}\sqrt{X|_{\omega_\text{LI}}^2+Y|_{\omega_\text{LI}}^2}\qquad\text{and}\qquad
    \tan(\varphi) = \frac{Y|_{\omega_\text{LI}}}{X|_{\omega_\text{LI}}}.
\eea
From the imparted change in the phase the signal delay is then calculated as:
\bea \nonumber
    \tau^{\text{(R)}} = \frac{\varphi}{\omega_\text{LI}}
\eea
where $\tau^{\text{(R)}}>0$ ($\tau^{\text{(R)}}<0$) represent a delay (advance) on the signal.

Here we have neglected the gain provided by the lock-in, which is important to determine the absolute value of $r(\omega_s)$. To account for that we calibrate the $X$ channel by a normalized transmission curve taken with low input power. Our assumption is that the cavity-taper coupling is not affected by the input power. A analogous result can be found for the case where the control laser is on the blue side of the cavity ($\omegap>\omega_o$). A more detailed description of the EIT experiment can be found in Ref.~\cite{Safavi-Naeini2011}.

\section{Analyzing the Mechanical Mode Spectra}
\label{AppG}
To determine the total spectral power at $\omega_m$ for a given measured spectra, we first subtract a background taken with the cooling laser far-detuned from the cavity (in the same calibration conditions). We then perform a least squares fit to a Lorentzian function of the form
\begin{equation}
L(\omega) = \frac{A}{\left(\frac{\omega-\omega_m}{2\gamma}\right)^2+1}
\end{equation}
with fit parameters $A$, $\omega_m$ and $\gamma$. The spectral power is then given simply by
\begin{equation}
P_{\omega_m} = \frac{A\gamma}{4}.
\end{equation}

To extract the intrinsic linewidth $\gamma_i$ we first fix the input power $P_\text{in}$. We then lock the pump on the red side of the cavity (at $\Delta = +\omega_m$) and measure the total linewidth, $\gamma_\text{red}=\gamma_i+\gamma_\text{OM}^\text{(red)}$. We repeat the measurement on the blue side (at $\Delta = -\omega_m$), where $\gamma_\text{blue}=\gamma_i-\gamma_\text{OM}^\text{(blue)}$. Using low input powers where $\gamma_\text{OM}^\text{(blue)}\ll\gamma_i$ to avoid amplification of the mechanical oscillations, we have $\gamma_\text{OM}^\text{(red)}=\gamma_\text{OM}^\text{(blue)}$, which leads to
\begin{equation}
\gamma_i = \frac{\gamma_\text{red} + \gamma_\text{blue} }{2}.
\end{equation}

\section{Measurement of Phonon Number and Error Analysis}
\label{AppH}
Equation~(\ref{eqn:PSB_out}) shows an explicit form for the sideband power amplitude seen by a photodetector for a red detuned pump laser. More specifically we can find a relation between the number of phonons inside the cavity and the power spectrum for our experimental setup. As shown before, the RF-spectra are detected via a RSA which displays the power spectral density of the voltage coming from the photodetector (with gain $G_e$ and $G_\text{EDFA}$). The single sided power spectral density at the detector is given by Equation~(\ref{eqn:PSB_out}), and denoted  $S_\text{PP}(\omega)$. Since the power is related to voltage by an electronic gain $G_e$, then $S_{VV}(\omega) = G_e^2 S_\text{PP}(\omega)$. When the EDFA is used, there is an additional gain factor, and $S_{VV}(\omega) = G^2_\text{EDFA} G_e^2 S_\text{PP}(\omega)$. Finally, the RSA reports power as opposed to squared voltages, and so the final spectral density measured is $S(\omega) = S_{VV}(\omega) / 2R_L$, where $R_L = 50~\Omega$ is the input impedance of the RSA and the factor of two in the denominator comes from the conversion of peak-to-peak voltage to RMS voltage. Then, in terms of integrated power, the power detected in the sideband on the RSA, $P_{\omega_m}$ is related to the heterodyne detected integrated spectral density by the relation
\begin{equation}
P_{\omega_m} = \frac{ (G_eG_\text{EDFA})^2}{2 R_L}P_\text{SB}.
\end{equation}
We would like to write this equation as a function of all the independent variables measured for the system, and from that estimate the error on the measured number of phonons.

From the DC transmission spectra, the optical components $\kappa$, $\kappa_e$ and $\omega_0$ can be determined. Both $G_\text{EDFA}$ and $G_e$ are measured and latter compensates for any discrepancy in the value of $R_L$. From the RF-spectra one can determine the total mechanical linewidth $\gamma = \gamma_i + \gamma_\text{OM}$, the mechanical frequency $\omega_m$ and the total RF-power $P_\text{RSA}$. The EIT spectra give the true detuning $\Delta$ between the pump laser and the cavity and, as shown before, can be used to determine the power at the cavity $P_\text{in}$.

Using Equations~(\ref{eqn:gamma_OM}) and (\ref{eqn:alpha_0}) we can rewrite the integrated form of Equation~(\ref{eqn:PSB_out}) as:
\begin{eqnarray}\label{eqn:PSB_out_modified}
P_\text{SB} = (\hbar\omega_0)^2 \frac{(\kappa_e/2) N_\text{in}}{(\Delta-\omega_m)^2+(\kappa/2)^2}~\kappa(\gamma - \gamma_i)\bar{n}
\end{eqnarray}

From Equation~(\ref{eqn:PSB_out_modified}) we can write the expression that relates the number of phonons, $\bar{n}$, and all the system parameters as:
\begin{eqnarray}\label{eqn:n_phonons}
\bar{n} = \left(\frac{2R_L}{G_e^2G_\text{EDFA}^2}\frac{P_{\omega_m}}{\hbar\omega_0} \right)\left( \frac{1}{\kappa(\gamma-\gamma_i)} \right)\left( \frac{(\Delta-\omega_m)^2+(\kappa/2)^2}{(\kappa_e/2)P_\text{in}} \right)
\end{eqnarray}

We can calculate the cumulative error for the number of measured phonons on the cavity using this equation and based upon on the measurable variables which gives:
\begin{eqnarray}\label{eqn:Dn_phonons}
\frac{\Delta\bar{n}}{\bar{n}} & = &\Bigg[ \frac{\delta\omega_0^2}{\omega_0^2}+
\frac{\delta\kappa_e^2}{\kappa_e^2} +
\frac{\delta P_\text{in}^2}{P_\text{in}^2} +
\frac{\delta P_\text{RSA}^2}{P_\text{RSA}^2}+
\frac{\delta\gamma_i^2}{(\gamma-\gamma_i)^2}+
\frac{\delta\gamma^2}{\gamma-\gamma_i)^2}+...\\\nonumber
& &+\left(\frac{\kappa/2}{(\kappa/2)^2+(\Delta-\omega_m)^2}-\frac{1}{\kappa}\right)^2\delta\kappa^2+ \left(\frac{2(\Delta-\omega_m)}{(\kappa/2)^2+(\Delta-\omega_m)^2}\right)^2\delta\Delta^2+
\left(\frac{2(\Delta-\omega_m)}{(\kappa/2)^2+(\Delta-\omega_m)^2}\right)^2\delta\omega_m^2
\Bigg]^{1/2}
\end{eqnarray}

Here we neglected the error on $G_e$ and $G_\text{EDFA}$ which are much smaller than any other error quantity. To determine the variation for $\kappa$, $\kappa_e$ and $\omega_0$, we measured the DC optical spectrum for every single data point in Figure~(4a) of the main text and determined $\delta\kappa$, $\delta\kappa_e$ and $\delta\omega_0$ from the normalized standard deviations of each of the values. The measurement uncertainty of these values are below $0.7\%$. The mechanical properties $\delta\gamma$, $\delta P_{\text{RSA}}$ and $\delta\omega_m$, were determined from the deviation on the spectra fits using a $95\%$ confidence interval, which produces percent errors below $0.6\%$. The pump laser detuning from the cavity is controlled by the EIT reflection spectra. To find the variation of the detuning $\delta\Delta$ we once again computed the standard deviation of all the measured detunings, which results in a deviation of less than $0.3\%$.

Finally, the two main sources of error in our data are the determination of the intrinsic mechanical quality factor (reflected in $\gamma_i$) and the input power, $P_{\text{in}}$. The uncertainty in the mechanical linewidth, $\delta\gamma_i$, is found by repeatedly measuring it at a single power level and computing its standard deviation (found to be $\sim1.6\%$). Using the calibration procedure discussed above for $P_{\text{in}}$, the error lies in the determination of losses $L_0$ and $L_1$. In the worst case the calibration would be off by the ratio between the input loss, $L_0$ in the present experiment, and the square root of total loss $\sqrt{L_0L_1}$ producing a percentage error of $\sim4.0\%$ to the input power.

Taking all of these factors into account produces an overall uncertainty of $\sim4.5\%$ in the measured absolute phonon number.

\section{Estimating the Temperature Shift from Thermo-optic Effects}
\label{AppI}

Absorption in the dielectric cavity causes the temperature of the dielectric cavity to increase locally. This effect is expressed through shifts in the refractive index of the structure, and the thermo-optic coefficient of Silicon~\cite{Frey2006}. As such, we can estimate the temperature of the cavity by looking at the shift in the cavity frequency, starting from a known temperature.

The starting point of the analysis is the cavity-perturbation formula for dielectric cavities~\cite{Harrington1961},
\begin{equation}
\frac{\omega-\omega_0}{\omega_0} \approx - \frac{1}{2}\frac{\int \delta \epsilon(\mathbf{r}) |\mathbf{E}(\mathbf{r})|^2 \mathrm{d}\mathbf{r}}{\int \epsilon(\mathbf{r}) |\mathbf{E}(\mathbf{r})|^2 \mathrm{d}\mathbf{r}}. \label{eqn:pertrubation}
\end{equation}

\begin{figure*}[ht!]
\begin{center}
\includegraphics[width=14cm]{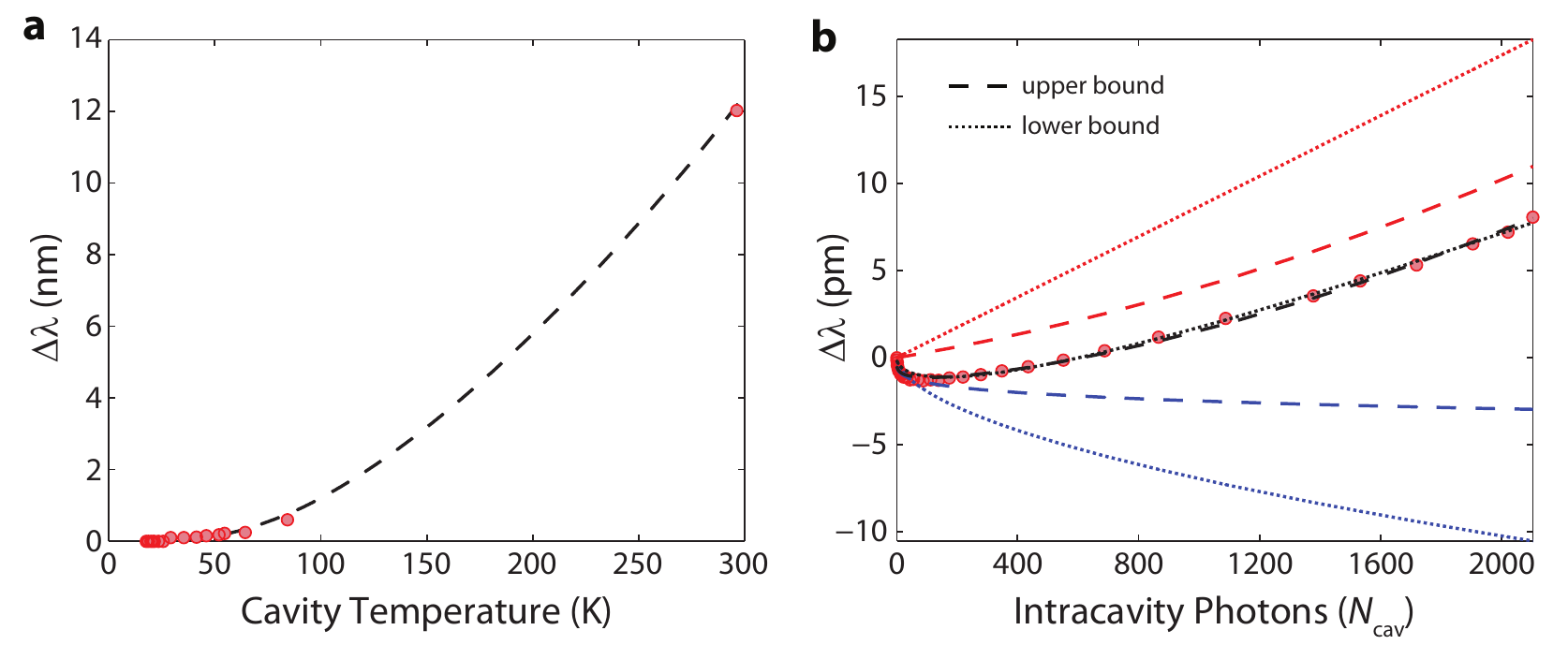}
\end{center}
\caption{\label{fig:thermooptic}\textbf{a}, the measured wavelength shift compared to the theoretical shift predicted by Equation (\ref{eqn:Tpertrubation}) for a range of cavity temperatures, using 17.6~K as the reference point. \textbf{b}, the measured power-dependent wavelength shift of the cavity with fitted individual contributions due to free carrier dispersion (blue) and refractive index change (red), as well as their sum (black), for the two bounds discussed in the text.}
\end{figure*}

From the relation $\epsilon/\epsilon_0 = n^2$, we find $\delta\epsilon = 2n\delta n \epsilon_0$. By assuming that the cavity as a whole is heated to a temperature $T_0$, the integral in Equation~(\ref{eqn:pertrubation}) can be written as
\begin{equation}
\omega-\omega_0 \approx -{ n(T_0)  \omega_0}\frac{\int_{\mathrm{Si}} |\mathbf{E}(\mathbf{r})|^2 \mathrm{d}\mathbf{r}}{\int (n(T_0))^2 |\mathbf{E}(\mathbf{r})|^2 \mathrm{d}\mathbf{r}}\times  (n(T) - n(T_0)). \label{eqn:Tpertrubation}
\end{equation}
Using the values of $n(T)$ found in literature~\cite{Frey2006}, and a value of $$\frac{\int_{\mathrm{Si}} |\mathbf{E}(\mathbf{r})|^2 \mathrm{d}\mathbf{r}}{\int (n(T_0))^2 |\mathbf{E}(\mathbf{r})|^2 \mathrm{d}\mathbf{r}}\approx 7.5066\times10^{-2},$$ calculated from the finite element simulations (FEM) of the mode profiles, we plot the wavelength shift from $17.6~\text{K}$ up to $300~\text{K}$ in Figure~(\ref{fig:thermooptic}a).
The total shift of $12.5~\text{nm}$ agrees with the experimentally observed change in resonance wavelength.

This analysis can be applied to the wavelength shift data for various input powers at low temperature where the initial cavity temperature is measured by thermometry methods discussed above. We note an initial blue-shift of the cavity, which is attributed to free-carrier dispersion effects~\cite{Barclay2005} and can be modeled by a power law dependence on intracavity photon number, $A(n_c)^B$. The temperature-dependent data for the refractive index of Silicon in Ref.~\cite{Frey2006} is valid only for $T>30$~K so the power-dependent cavity heating for a starting temperature of $17.6$~K, for the largest intracavity photon number, can only be bounded. For the upper bound, we assume $dn/dT = 0$ for $T<30$~K, resulting in a $\Delta T_\text{max}$ of $16.8$~K. For the lower bound, we assume $dn/dT = dn/dT|_{T=30~\text{K}}$ for $T<30$~K, resulting in a $\Delta T_\text{min}$ of 7.8~K. These bounds and their respective fits are shown in Figure~(\ref{fig:thermooptic}b).

\section{Temperature-Dependent modifications to the intrinsic mechanical damping}
\label{AppJ}
\begin{figure*}[ht!]
\begin{center}
\includegraphics[width=14cm]{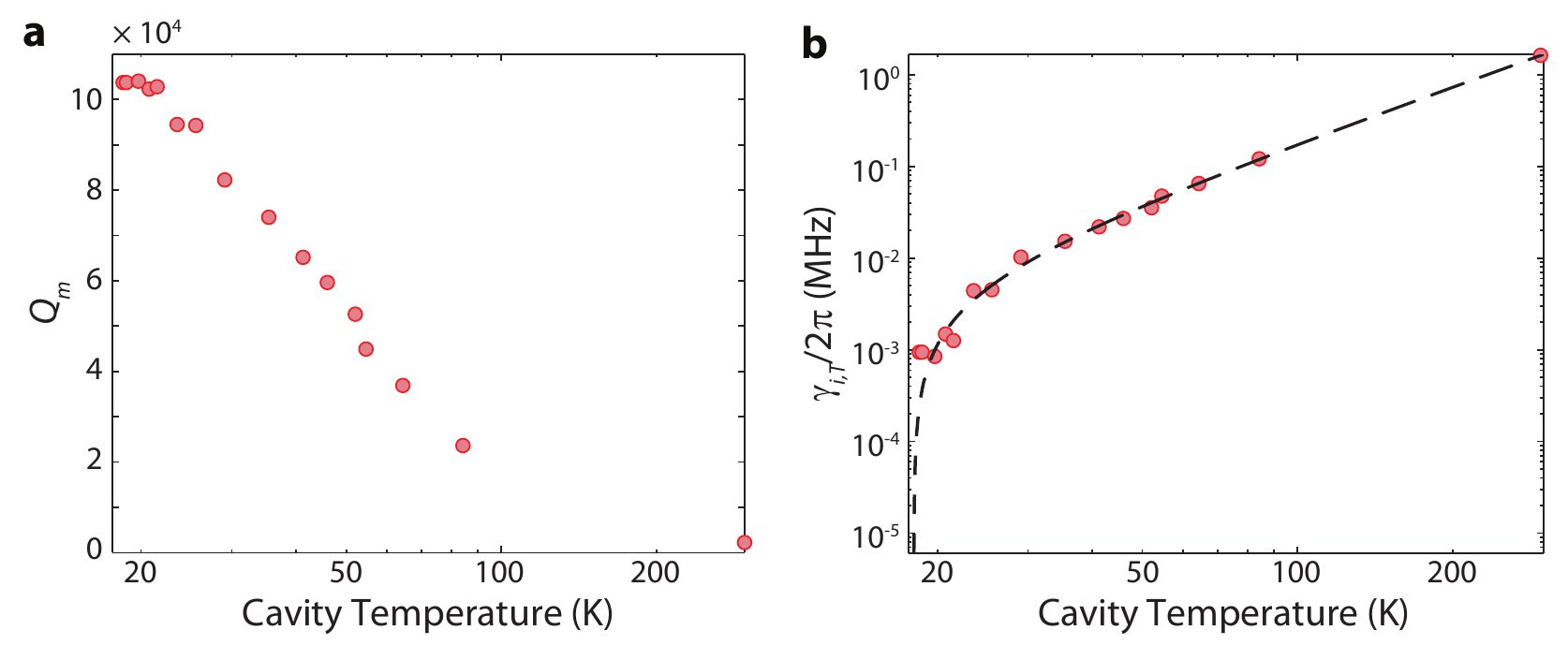}
\end{center}
\caption{\label{fig:gammait}\textbf{a}, the measured intrinsic mechanical quality factor for various cavity temperatures. \textbf{b}, the inferred intrinsic mechanical loss rate due to temperature, $\gamma_{i,T}$, modeled by a polynomial fit.}
\end{figure*}
Independent measurements of the mechanical quality factor, $Q_m$, at varying bath temperatures indicate that the $Q_m$ changes with temperature (Figure~(\ref{fig:gammait}a)). These measurements are taken at low intracavity photon number, rendering free-carrier effects negligible. As such we can model the mechanical loss rate as $\gamma_i(T) = \gamma_{i,T} (T) + \gamma_i^{(0)}$ where $\gamma_i^{(0)}$ is the measured loss rate at the reference temperature (17.6~K). The extracted form of $\gamma_{i,T}(T)$ is shown in Figure~(\ref{fig:gammait}b).

\section{Photon-number dependent modifications to the intrinsic mechanical damping}
\label{AppK}

The deviation of the expected cooled phonon number from the measured value is a result of two factors: bath heating and an increase in the intrinsic mechanical loss rate $(\gamma_i)$ due to heating and free carriers. Since the integrated spectral power of the mechanical mode depends only on the product $\gamma_i T_b$ (for large intracavity photon numbers $n_c$), naively ignoring the latter effect results in an estimated change of $\Delta T>50$~K in the bath temperature for 2000 intracavity photons. This is unrealistic as such a temperature change would tune the optical mode red by $>300$~pm (from Equation~(\ref{eqn:Tpertrubation})), while the actual measured shift is closer to $10-20$~pm (from Figure~(\ref{fig:thermooptic}b)). In fact through independent measurements (where dynamic back-action was minimized), we found that the mechanical linewidth is a function of the number of photons in the cavity. We attribute this to a nonlinear process in the cavity involving the generation of free carriers, which will be explored in depth elsewhere, and introduce an additional loss channel $\gamma_{i,\text{FC}}$ in the mechanical loss rate so that we have
\begin{equation}
\gamma_i \rightarrow \gamma_i \equiv \gamma_i^{(0)} + \gamma_{i,T}(T(n_c)) + \gamma_{i,\text{FC}}(n_c).\label{eqn:gammaitotal}
\end{equation}

\begin{figure*}[ht!]
\begin{center}
\includegraphics[width=14cm]{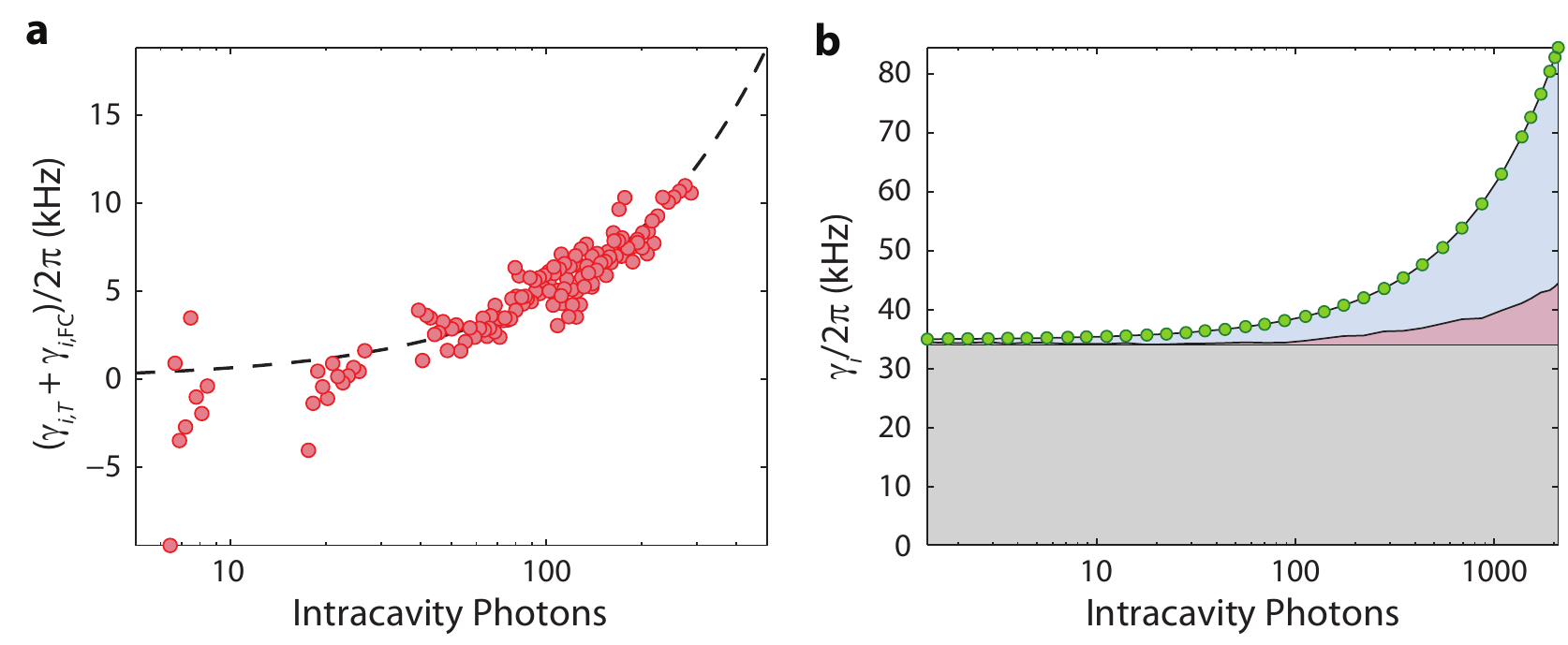}
\end{center}
\caption{\label{fig:gammaitotal} \textbf{a}, excess loss as a function of $n_c$, inferred from far-red-detuned ($\Delta > 5.5$~GHz) measurements of $\gamma_\text{cooled}$. \textbf{b}, breakdown showing the individual contributions of $\gamma_i^{(0)}$ (gray), $\gamma_{i,T}$ (red), and $\gamma_{i,\text{FC}}$ (blue) to the total $\gamma_i$ ($\circ$). }
\end{figure*}

From the relations shown on previous sections, we have $\gamma_\text{cooled}^{(0)} = \gamma_i^{(0)} + \gamma_\text{OM}$. Incorporating Equation~(\ref{eqn:gammaitotal}), we have experimentally, $\gamma_\text{cooled} = \gamma_i + \gamma_\text{OM}$, with their difference yielding the magnitude of the additional loss rates. However, for  $\Delta = \omega_m$ and $n_c>10$, $\gamma_\text{OM}$ tends to be large compared to $\gamma_i$, making this subtraction quite error prone. Thus, to get accurate data for high intracavity photon numbers we use a range of larger detunings, noting that $\gamma_\text{OM} \propto \Delta^{-2}$ for $\Delta \gg \omega_m$ and fixed $n_c$ (approximately). This loss is then modeled using a power law dependence on $n_c$ (Figure~(\ref{fig:gammaitotal}a)).

Using the models of mechanical loss from Figure~(\ref{fig:gammaitotal}a) and Figure~(\ref{fig:gammait}b) with the thermometry technique outlined earlier (making the replacement to $\gamma_i$) allows a more accurate determination of the temperature rise in the cavity, as well as the characterization of the individual contributions of $\gamma_{i,T}$ and $\gamma_{i,\text{FC}}$ as a function of $n_c$. The result is an estimated increase of 13.2~K in $T_b$ at the highest input power, well within the previously fitted temperature bounds.

\begin{figure*}[ht!]
\begin{center}
\includegraphics[width=14cm]{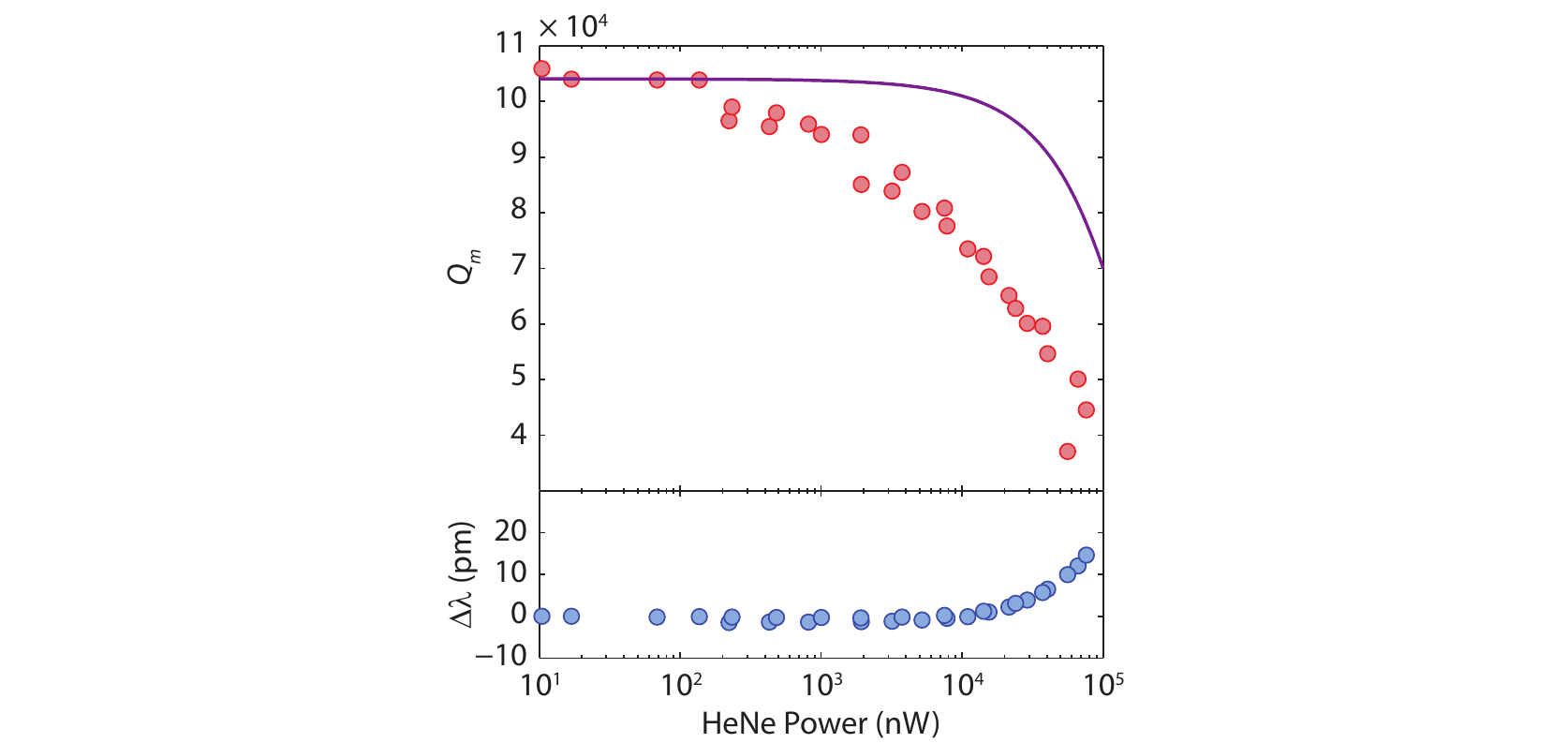}
\end{center}
\caption{\label{fig:greenlaser} The $Q_m$ degradation as a function of $532$~nm laser power. The purple line shows the expected $Q_m$ for the bath temperature rise inferred from the wavelength shift data. The deviation of the data from this prediction suggests an additional loss channel related to the presence of free carriers. }
\end{figure*}

The addition of a free carrier related loss channel is further corroborated by pumping the Si sample above the band gap with a $532$~nm solid state green laser, directly stimulating the production of free carriers. The degradation in $Q_m$ can be only partially explained by heating due to absorption since the maximum $19$~K temperature rise estimated from the cavity red-shift results in an expected $Q_m$ of approximately 70,000 at the highest power (Figure~(\ref{fig:gammait}a)), whereas a far lower value is measured. The remaining excess loss is attributed to the presence of free-carriers. These results are shown in Figure~(\ref{fig:greenlaser}).

\section{Shot Noise Considerations}
\label{AppL}
\begin{figure*}[ht!]
\begin{center}
\includegraphics[width=14cm]{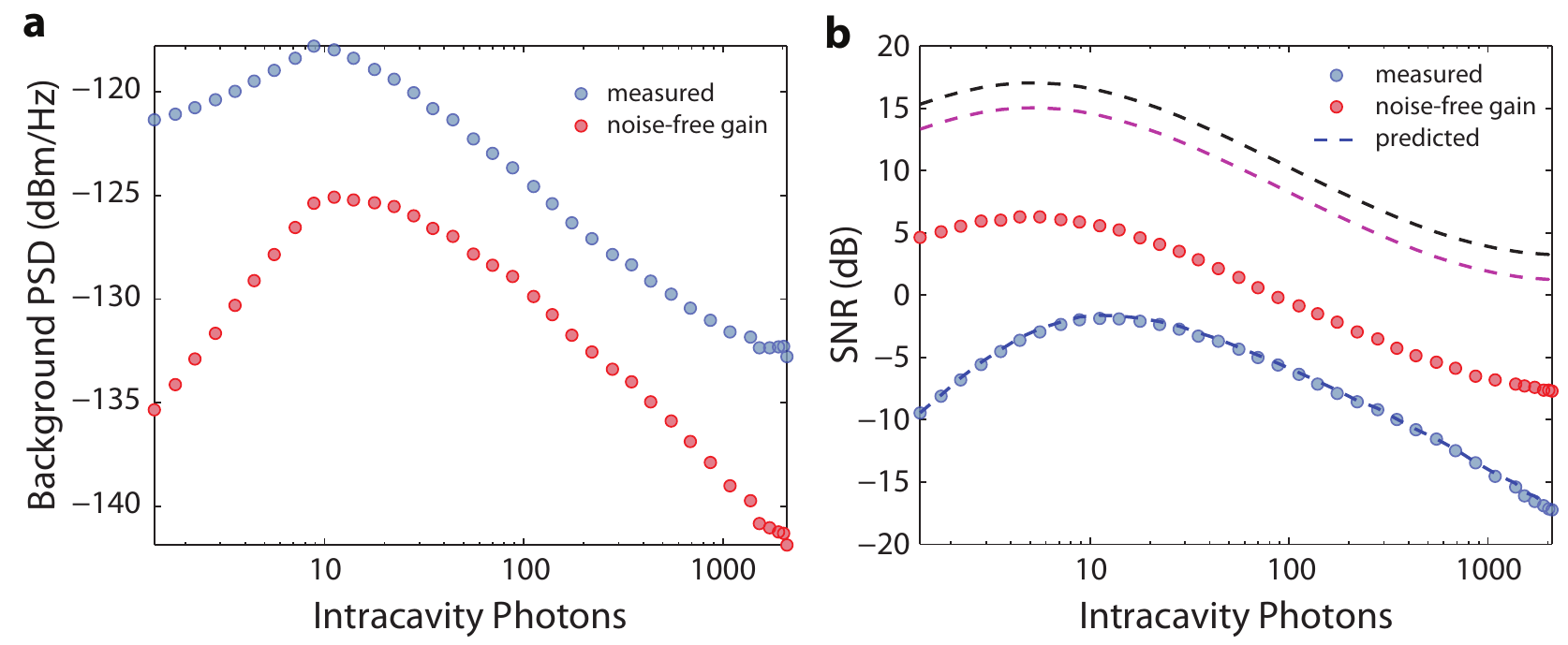}
\end{center}
\caption{\label{fig:sn} \textbf{a}, a comparison of the measured background spectrum against the shot noise level amplified by an ideal, noise-free amplifier. \textbf{b}, a comparison of the measured signal-to noise ratio (SNR) to the maximum achievable SNR for the experimental setup assuming an ideal, noise-free amplifier and perfect quantum detection; the predicted SNR uses only measured device and calibration parameters, along with $S^2_\text{excess}$ determined from (a); the purple dashed line shows SNR assuming an overcoupled system, where $\kappa_e = \kappa$, with noise-free gain and the black line shows the same system without taper/insertion loss (representing the most ideal of ideal cases).}
\end{figure*}
We consider here the impact of using a non-ideal amplifier (EDFA) on the measured signal, and the deviation from quantum limits. For a coherent optical beam with frequency $\omegap$ and power $P$ incident on a photo detector, the single sided power spectral density of the shot noise is simply
\begin{equation}
S_\text{shot}(\omega) = \sqrt{2\hbar\omegap P},\label{eqn:shot}
\end{equation}
independent of frequency. As part of the calibration procedure, $G_\text{EDFA}$ is measured to characterize the gain provided by the EDFA and $G_e$ is measured to characterize the transimpedance gain and quantum efficiency of the photodetector. These values can be used in conjunction with Equation~(\ref{eqn:shot}) to predict the expected spectral background assuming only the presence of shot noise and noise-free gain. To wit,
\begin{equation}
S_\text{shot}^\text{(amplified)} = \sqrt{2\hbar\omega_oG_{\text{EDFA}}^2G_e^2P_{\text{in}}'},
\end{equation}
where the prime indicates the value has been adjusted to account for insertion loss from the cavity to the photodetector. The difference between this predicted level and the measured background level, $S_\text{background}$, gives the non-ideality of the EDFA and is attributed to an excess noise which also includes the amplified spontaneous emission (ASE) noise~\cite{Desurvire2002}. This deviation is shown in Figure~(\ref{fig:sn}a). Defining
\begin{equation}
S^2_\text{excess} = S^2_\text{background} - (S^\text{(amplified)}_\text{shot})^2
\end{equation}
this additional noise reduces the measured signal-to-noise ratio (SNR). We can predict the shot noise limited SNR, representing the largest measurable SNR for the current experimental setup set assuming perfectly efficient detection from the measured device and calibration parameters using Equation (\ref{eqn:PSB_out_modified}) and (\ref{eqn:shot}). This ratio is given by
\begin{equation}
\mbox{SNR}_\text{shot} = \frac{{4 P_\text{SB}' }/{(\gamma_i+\gamma_\text{OM})}}{2\hbar\omega_o P_\text{in}'}.
\end{equation}
Similarly, the expected measurement SNR, using only device and calibration parameters, is given by
\begin{equation}
\mbox{SNR}_\text{predicted} = \frac{{4 G_{\text{EDFA}}^2G_e^2 P_\text{SB}' }/{(\gamma_i+\gamma_\text{OM})}}{2  \hbar\omega_oG_\text{EDFA}^2G_e^2 P_\text{in}' + S_\text{excess}^2},
\end{equation}
which closely corresponds to the measured values, as seen in Figure~(\ref{fig:sn}b).

\end{document}